\documentclass[prb,superscriptaddress,floatfix,letterpaper,twocolumn,aps,showpacs]{revtex4-1}
\usepackage[utf8]{inputenc}
\usepackage{natbib}
\usepackage{amsmath}
\usepackage{amsbsy}
\usepackage{amssymb}
\usepackage{scrextend}
\usepackage{graphicx}
\usepackage{color}
\usepackage{xcolor}

\newcommand{\fig}{Fig.\ }
\newcommand{\eq}{Eq.\ }
\newcommand{\eqs}{Eqs.\ }
\newcommand{\sect}{Sec.\ }

\newcommand{\sbonlinecite}[1]{[\onlinecite{#1}]}

\newcommand{\C}{\bar{C}}

\newcommand{\epin}{e_{\mathrm{pin}}}
\renewcommand{\FL}{F_{\mathrm{\scriptscriptstyle L}}}
\newcommand{\Fpin}{F_{\mathrm{pin}}}
\newcommand{\fpin}{f_{\mathrm{pin}}}

\renewcommand{\vec}[1]{\boldsymbol{#1}}
\newcommand{\av}[1]{\langle #1 \rangle}

\begin{document}
\title{Probing the pinning landscape in type-II superconductors via Campbell
penetration depth}
\author{R.\ Willa}
\author{V.B.\ Geshkenbein}
\author{G.\ Blatter}
\affiliation{Institute for Theoretical Physics, ETH Zurich, 8093 Zurich,
Switzerland}
\date{\today}

\begin{abstract}
Type-II superconductors owe their magnetic and transport properties to vortex
pinning, the immobilization of flux quanta through material inhomogeneities or
defects. Characterizing the potential energy landscape for vortices, the
pinning landscape (or short, pinscape), is of great technological importance.
Besides measurement of the critical current density $j_c$ and of creep rates
$S$, the $ac$ magnetic response provides valuable information on the pinscape
which is different from that obtained through $j_c$ or $S$, with the Campbell
penetration depth $\lambda_{\rm \scriptscriptstyle C}$ defining a
characteristic quantity well accessible in an experiment. Here, we derive a
microscopic expression for the Campbell penetration depth $\lambda_{\rm
\scriptscriptstyle C}$ using strong pinning theory. Our results explain the
dependence of $\lambda_{\rm \scriptscriptstyle C}$ on the state preparation of
the vortex system and the appearance of hysteretic response. Analyzing
different pinning models, metallic or insulating inclusions as well as $\delta
T_c$- and $\delta \ell$-pinning, we discuss the behavior of the Campbell
length for different vortex state preparations within the phenomenological
$H$-$T$ phase diagram and compare our results with recent experiments.
\end{abstract}

\maketitle
\section{Introduction}\label{sec:intro}
In a type-II superconductor, the magnetic field $H$ penetrates the material in
the form of vortices \cite{Shubnikov1937,Abrikosov1957}, individually
capturing a superconducting flux quantum $\Phi_{0} = hc/2e$ and together
forming a triangular Abrikosov lattice generating the magnetic induction $B$
inside the sample. In ideal superconductors, an applied current density $j$
generates a Lorentz force $F_{\mathrm{\scriptscriptstyle L}} = j B/c$, setting
the vortex lattice in motion. The resulting velocity $v$ produces an electric
field $E = vB/c$ which renders the current transport dissipative
\cite{Bardeen1965}. The material's response then is characterized by the
flux-flow resistivity $\rho_\mathrm{ff} \simeq \rho_n B/H_{c2}$, with $\rho_n$
the normal state resistivity and $H_{c2}$ the upper critical field.  In real
materials, chemical impurities or crystallographic defects immobilize
vortices, thereby restoring dissipation-free electric transport for currents
$j$ below a critical value $j_c$. Achieving large critical currents $j_c$ is a
prime task in optimizing superconductors for technological applications.
Recently, analytic, numerical, and experimental studies have been used in a
concerted effort to investigate the fundamental mechanisms governing vortex
pinning \cite{Sadovskyy2015a,Sadovskyy2015b}. Such a program relies on a
proper characterization of the material's pinning landscape or pinscape.
Besides measurement of the critical current density $j_{c}$, the analysis of
the material's $ac$ magnetic response\cite{Campbell1969,Campbell1971} as
quantified through the Campbell penetration depth $\lambda_{\mathrm{
\scriptscriptstyle C}}$ provides valuable infomation on the bulk pinning
parameters. In this paper, we present a microscopic foundation for the
Campbell penetration length $\lambda_{\mathrm{ \scriptscriptstyle C}}$ which
allows to connect the result of $ac$ magnetic response measurements to
microscopic parameters of the pinscape.

When measuring a material's $ac$ response, a small magnetic field $h_{ac}$ is
applied on top of a large $dc$ field $B_0$.  In Campbell's original
phenomenological approach \cite{Campbell1969,Campbell1971}, the $ac$ field
forces vortices to oscillate within their pinning potentials which are
conveniently characterized by an effective harmonic potential well $\alpha
u^{2}/2$, with $u$ denoting the vortex displacement. A measurement of the
Campbell length $\lambda_{\mathrm{\scriptscriptstyle C}} \propto
\alpha^{-1/2}$ then informs about the curvature of this 'single-vortex'
potential. Later, the $ac$ magnetic response has been further discussed in the
context toy models\cite{Lowell1972_1,Campbell1978_1} assuming a piecewise
parabolic potentials. In the wake of the discovery of high-temperature
superconductivity \cite{Bednorz1986}, the frequency response of the vortex
state has attracted renewed attention, especially in the context of vortex
creep \cite{Brandt1991,Koshelev1991,Coffey1991,Coffey1992,vanderBeek1993}.
Recent experimental developments in the field have been reviewed in Refs.\
\sbonlinecite{Gomory1997} and \sbonlinecite{Prozorov2003}.

A phenomenological approach as described above cannot relate the measured
penetration depth $\lambda_{\mathrm{\scriptscriptstyle C}}$ to the microscopic
parameters of the pinscape. In particular, it is unclear how such a simple
description can deal with the Bean critical state\cite{Bean1962}. The latter
is realized at $j = j_{c}$ where the pinning landscape acts with
its maximal force $F_c$ against the Lorentz force $F_{\mathrm{
\scriptscriptstyle L}}$ and establishes a self-organized critical state
resembling a sandpile \cite{Bak1987}, with avalanche-type motion of vortices
triggered upon increasing the magnetic field. The phenomenological
model\cite{Prozorov2003} describes this situation by a vanishing
curvature $\alpha(j \to j_{c}) \to 0$, resulting in a formally diverging
Campbell length and hence a full penetration of the $ac$ signal. Such a
divergent signature has not been observed in experiments\cite{Prozorov2003};
rather, it has been found that the Campbell length can even decrease when
going from a field-cooled state (FC) to a Bean critical (or zero-field-cooled,
ZFC) state \cite{Kim2013}.

Vortex pinning, including $j_c$, is usually described within either of two
frameworks, weak collective pinning due to the joint action of many weak
defects or strong pinning produced by a low density of strong impurities
\cite{Larkin1979,Blatter2004}.  Within our microscopic description, we make
use of strong pinning theory and relate the measured penetration depth
$\lambda_{\mathrm{\scriptscriptstyle C}}$ to microscopic parameters of the
pinscape. Most interestingly, it turns out that $j_c$ and
$\lambda_{\mathrm{\scriptscriptstyle C}}$ are determined by different
microscopic parameters: while the critical current density $j_c$ involves the
characteristic jump in energy of strong pinning theory, the Campbell length
involves the jump in the pinning force. The scaling $j_c \sim c\xi B
/\lambda_{\mathrm{\scriptscriptstyle C}}^2$ used in the past then picks up a
non-trivial dependence on the strong pinning (or Labusch) parameter
$\kappa>1$, e.g., $j_c \sim (c \xi B / \lambda_{\mathrm{ \scriptscriptstyle
C}}^2) (\kappa - 1)^{3/2}$ at the onset of strong pinning when $\kappa - 1 \ll
1$ and $j_c \sim (c \xi B / \lambda_{\mathrm{\scriptscriptstyle C}}^2) \kappa$
for very strong pinning $\kappa \gg 1$. The quantitative power of strong
pinning theory provides further interesting results such as the dependence of
$\lambda_{\mathrm{\scriptscriptstyle C}}$ on the vortex state (e.g.,
field-cooled versus zero-field-cooled) or the appearance of hysteretic
behavior upon temperature cycling. Furthermore, mapping out the behavior of
$\lambda_{\mathrm{\scriptscriptstyle C}}$ within the $H$-$T$ phase diagram
allows to draw interesting conclusions on the character of the pinning
centers. While our analysis focuses on bulk characteristic parameters of the
pinscape, different types of scanning techniques have been used recently to
obtain a direct microscopic image of the potential landscape seen by
individual vortices\cite{Timmermans2014,Embon2015}.

In the following, we first review the general approach to the $ac$ response
(\sect \ref{sec:macroscopic-ac-response}) and then introduce the strong
pinning formalism in \sect \ref{sec:strong-pinning}.  We derive a quantitative
relation between the Campbell length and the microscopic pinning potential and
discuss the generic dependence of $\lambda_{\mathrm{ \scriptscriptstyle C}}$
on the state preparation (e.g., FC vs.\ ZFC) as well as hysteretic effects. In
\sect \ref{sec:specific-pinning-models}, we analyze the pinning
characteristics of four types of defects, namely insulating and metallic
inclusions as well as $\delta T_{c}$- and $\delta \ell$-pins. In \sect
\ref{sec:experiments} we compare our findings with recent measurements
\cite{Kim2013} on SrPd$_{2}$Ge$_{2}$ and find good overall agreement using a
pinscape with a scaling characteristic of insulating defects.  A brief account
of parts of this work can be found in Refs.\ \sbonlinecite{Willa2015a} and
\sbonlinecite{Willa2015b}.

\section{$ac$ magnetic response}\label{sec:macroscopic-ac-response}
We analyze the magnetic response of a bulk superconducting sample subject to a
static field $H$ and a parallel $ac$ perturbation with a small amplitude
$h_{ac} \ll H$. While the magnetic field $H$ induces a vortex lattice with an
average induction $B_0$ in the sample, the small $ac$-field induces motion of
these vortices. We choose a geometry with a superconductor filling the
half-space $X > 0$ with the sample surface, magnetic field, and $Z$-axis
arranged in parallel. This corresponds, up to finite size effects, to a sample
in a slab geometry arranged parallel to $Z$ where demagnetization effects are
absent. We will briefly discuss the geometry of a thin platelet-like sample
(arranged in the $XY$-plane) at the end of the section.

On the \emph{macroscopic} level, the vortex lattice can be described as an
elastic medium and its response to the $ac$ perturbation is reflected in a
\emph{macroscopic} displacement field $U(X,t)$ of the flux-lines. We use
capital letters when describing macroscopic coordinates and displacements and
denote their microscopic counterparts (below) by lower-case symbols.  Starting
from an initial field $B_{0}(X)$ and current $j_{0}(X)$ at time $t=0$, the
vector potential $\delta A = U(X,t) B_{0}(X)$ associated with the vortex
lattice displacement $U$ induces time-dependent corrections of the form
\begin{align}\label{eq:deltaB}
   \delta B (X,t) &= - B_{0}(0) \partial_{X} U(X,t),\\
   \label{eq:deltaj}
   \delta j (X,t) &= \frac{c}{4\pi} B_{0}(0)\partial^{2}_{X}U(X,t).
\end{align}
The above expressions are valid in the linear-response regime where $U$ is the
smallest length and $(\partial_{X}B_{0})U \ll B_{0}(\partial_{X}U)$.
Integrating \eq \eqref{eq:deltaB} over $X$, we find the flux $\phi(t)$ (per
unit of length along $y$) that has penetrated the surface,
\begin{align}\label{}
   \phi(t) = \int_{0}^{\infty} dX\, \delta B(X,t) = B_{0} U(0,t).
\end{align}
The distribution of this flux within the sample generates an additional
induction $\delta B(X,t)$ on top of the $dc$ field $B_0(X)$, see also Sec.\
\ref{sec:campbell-length}.

Having reduced the change in fields and currents to the macroscopic
displacement field $U(X,t)$ of vortices, we can find the dynamical response of
$B(X,t)$ through the equation of motion of the flux-line lattice,
\begin{align}\label{eq:eq-of-motion}
   \eta \dot{U} &= \FL(j,U) + \Fpin(X, U),
\end{align}
which balances the dissipative Bardeen-Stephen term with $\eta = B
H_{c2}/\rho_{n} c^{2}$ against the sum of Lorentz and pinning force densities,
$\FL = (j_{0} + \delta j) B_{0}/c$ and $\Fpin = F_{0} + \delta \Fpin$. The
static initial state is characterized by a pinning force $F_{0}$ that exactly
compensates the Lorentz force $j_{0}B_{0}/c$ and the right-hand side vanishes
identically (for a field-cooled sample both $F_0$ and $j_0$ vanish
individually).  Hence, the dynamic equation \eqref{eq:eq-of-motion} assumes
the form
\begin{align}\label{eq:eom}
   \eta \dot{U} - \frac{B_{0}^{2}}{4\pi} \partial_{X}^{2} U 
   - \delta \Fpin(X, U) = 0.
\end{align}
Making use of \eq \eqref{eq:deltaB}, the external drive $\delta B (X = 0,t) =
h_{ac} e^{-i \omega t}$ determines the boundary condition $\partial_{X}U(X,t)
= - (h_{ac}/B_{0}) e^{-i \omega t}$. It remains to find an expression for the
change in pinning force density $\delta \Fpin(U)$. 

Referring to Campbell's original work \cite{Campbell1969,Campbell1971}, one
usually assumes that vortices oscillate reversibly in an effective parabolic
pinning potential $\alpha U^{2}/2$, what results in the phenomenological
pinning force density
\begin{align}\label{eq:phenom-restoring-force}
   \delta \Fpin(U) = - \alpha\, U.
\end{align}
Using this Ansatz, the equation of motion \eqref{eq:eom} can readily be solved
for the displacement field $U(X,t)$, from which the field and current dynamics
follow via Eqs.\ \eqref{eq:deltaB} and \eqref{eq:deltaj}. One finds that the
field oscillations
\begin{align}\label{}
   \delta B(X,t) &= h_{ac} e^{-X/\lambda_{\mathrm{\scriptscriptstyle C}}} 
   e^{-i \omega t}
\end{align}
decay into the sample with the characteristic length $\lambda_{\mathrm{
\scriptscriptstyle C}}(\omega)$, which reduces to the Campbell length
\begin{align}\label{eq:lC}
   \lambda_{\mathrm{\scriptscriptstyle C}}
   = \Bigl(\frac{B_{0}^{2}}{4\pi \alpha}\Bigr)^{1/2}
\end{align}
at low frequencies $\omega \to 0$. The Campbell length thus relates the $ac$
penetration depth with the pinning properties of vortices through the
curvature $\alpha$ of the effective pinning potential, a relation that has
widely been used to characterize the pinning landscape.

However, inferring the properties of the pinning landscape from the
experimental measurements based on a simple phenomenological model is prone to
misjudgments. We therefore proceed with a \emph{microscopic} approach based on
strong pinning theory\cite{Labusch1969,Larkin1979} in order to determine the
\emph{macroscopic} response of the vortex state.  Thereby the macroscopic
equation of motion Eq.\ \eqref{eq:eom} can equally well be obtained from a
microscopic route by averaging the equation of motion of individual vortices
over an area much larger than $a_{0}^{2}$, with $a_{0}$ the inter-vortex
distance. The evaluation of the change in pinning force $\delta \Fpin(X, U)$,
involves a proper average of the microscopic action of single pinning centers,
a task we address in the following.

Before doing so, we briefly touch upon geometric aspects of the problem.  For
the slab geometry chosen here (with magnetic fields along $Z$ and currents
along $Y$), the contributions from shear and tilt deformations average to zero
and only the bulk compression modulus $c_{11}(\vec{k} = 0) = B_{0}^{2}/4\pi$
enters in \eq \eqref{eq:eom}.  In the platelet geometry, as opposed to the
slab geometry, the field is arranged perpendicular to the largest sample
dimension and demagnetization effects change both the size and effective
direction of the $ac$ field component.  For a sample thicker than $2
\lambda_{\mathrm{\scriptscriptstyle C}}$, the $ac$-component is screened and
the effective drive $h_{ac}^\mathrm{eff}$ is enhanced by $(w/d)^{1/2}$ and
redirected parallel to the surface (here, $w$ and $d$ denote the width and
thickness of the sample), see e.g., Refs.\ \sbonlinecite{Zeldov1994} and
\sbonlinecite{Willa2014}. The penetration of the $ac$-field then corresponds
to tilting the vortices within a depth $\lambda_{\mathrm{ \scriptscriptstyle
C}}$ away from the surface and the relevant distortion modulus appearing in
\eq \eqref{eq:eom} is the bulk tilt modulus $c_{44}(\vec{k} = 0) =
B_0^2/4\pi$.

\section{Strong pinning}\label{sec:strong-pinning}
\subsection{Formalism}\label{sec:formalism}
Within strong pinning theory as originally discussed by Labusch
\cite{Labusch1969} and later by Larkin and Ovchinnikov \cite{Larkin1979}, a
low density $n_{p}$ of pinning sites produces a finite pinning force by
inducing large plastic deformations on the pinned vortices. We consider a
lattice of vortices (directed along $z$) with equilibrium coordinates
$\vec{r}_{\mu} = (x_{\mu}, y_{\mu})$ and an isolated defect at the origin
defined through its pinning potential $e_{p}(\vec{r},z) \simeq
e_{p}(\vec{r}) \delta(z)$, with $\vec{r}=(x,y)$; as pins act independently,
the action of a finite density of pins is trivially summed over. The
interaction of the pin with the vortex lattice gains the system a local
energy density
\begin{align}\label{eq:energy-of-single-pin}
   \varepsilon_{p}(\vec{r},z;\vec{u})
        = \sum\limits_{\mu} e_{p}(\vec{r})\delta(z)\,
          \delta^{2}\{\vec{r}-[\vec{r}_{\mu} + \vec{u}(\vec{r}_{\mu},z)]\},
\end{align}
where $\vec{r}_{\mu} +\vec{u}(\vec{r}_{\mu},z)$ is the real position of the
$\mu$-th flux line with $\vec{u}(\vec{r}_{\mu},z)$ its \emph{microscopic}
displacement field away from the equilibrium position $\vec{r}_{\mu}$.
Variational minimization of the elastic deformation and pinning energies
results \cite{Blatter1994,Larkin1979} in an inhomogeneous differential
equation; its solution can formally be expressed through a self-consistency
condition involving the lattice's elastic Green's function $G_{\alpha
\beta}(\vec{r},z)$,
\begin{align}\label{eq:self-consistency-eq}
   u_{\alpha}(\vec{r}_{\nu},z_{\nu}) &= \\ \nonumber
      \int & dz\, d^{2}r\, G_{\alpha \beta}(\vec{r}_{\nu} - \vec{r}, z_{\nu} - z) 
   \big[\!-\!\partial_{u_{\beta}} \varepsilon_{p}(\vec{r},z,\vec{u})\big].
\end{align}
Here, $\alpha$ and $\beta$ index the in-plane components $x$ and $y$, and
$\nu$ is a vortex label and we assume summation over double indices.
Inserting \eq \eqref{eq:energy-of-single-pin} into \eq
\eqref{eq:self-consistency-eq} and defining the pinning force profile
$\vec{f}_{\!p}(\vec{r}) = -\nabla_{\!\vec{r}}\, e_{p}(\vec{r})$, we find to
dominant order in $\vec{u}$
\begin{align}
   u_{\alpha}(\vec{r}_{\nu},z_{\nu}) = \sum_{\mu}
      G_{\alpha \beta}(\vec{r}_{\nu} - \vec{r}_{\mu}, z_{\nu})
      f_{p,\beta}[\vec{r}_{\mu} + \vec{u}(\vec{r}_{\mu},0)].
\end{align}%
\begin{figure}[tb]
\includegraphics[width=6.0cm]{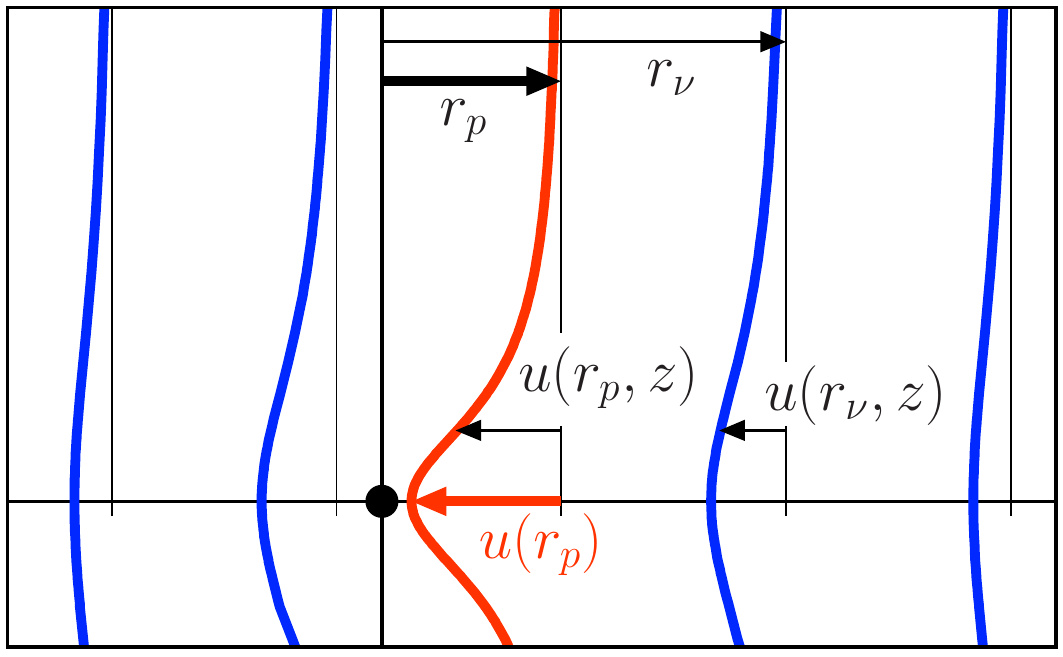}
\caption{Schematic view of the vortex distortion near a pinning center. The
pinned vortex (red) with asymptotic position $\vec{r}_{\mu} = \vec{r}_{p}$ is
deformed to $\vec{r}_{p} + \vec{u}(\vec{r}_{p},z)$ by the presence of the pin
(black dot). Nearby vortices with $\nu \neq \mu$ (blue) are deformed as well
with their deformation $\vec{u}(\vec{r}_{\nu},z)$ transported by the elastic
Green's function $G(\vec{r}_{\nu}-\vec{r}_{p},z_{\nu}-z)$. The full pinning
problem can be reduced to a self-consistency equation for the deformation
$\vec{u}(\vec{r}_{p})\equiv \vec{u}(\vec{r}_{p},0)$ of the pinned vortex at
the height $z=0$ of the defect.}
\label{fig:vortex-parameters}
\end{figure}
For a pinning potential with a trapping range smaller than the inter-vortex
distance $a_{0}$ and pinning at most one vortex, only one term is relevant in
the above summation and we arrive at
\begin{align}\label{eq:self-consistency-eq-3}
   u_{\alpha}(\vec{r}_{\nu},z_{\nu}) =
      G_{\alpha \beta}(\vec{r}_{\nu} - \vec{r}_{p}, z_{\nu})
      f_{p,\beta}[\vec{r}_{p} + \vec{u}(\vec{r}_{p},0)],
\end{align}
with $\vec{r}_{p} = \vec{r}_{\mu}$ the equilibrium position of the vortex in
the vicinity of the pinning site. Evaluating \eqref{eq:self-consistency-eq-3}
for this vortex at $\vec{r}_{\nu} = \vec{r}_{p}$ and $z_{\nu} = 0$, we obtain
the self-consistency condition
\begin{align}\label{eq:self-consistency-eq-4}
   u_{\alpha}(\vec{r}_{p}) &= 
      G_{\alpha \beta}(\vec{0}, 0) f_{p,\beta}[\vec{r}_{p} + \vec{u}(\vec{r}_{p})]
\end{align}
for the displacement $u_{\alpha}(\vec{r}_{p}) \equiv u_{\alpha}
(\vec{r}_{p},0)$ of the vortex pinned at the defect site. This expression can
be further simplified by exploiting the isotropy of the local Green's
function, $G_{\alpha \beta}(\vec{0}, 0) = \delta_{\alpha \beta} / \bar{C}$,
with the effective elasticity $\C$ defined through
\begin{align}\label{eq:Cbar-def}
   \bar{C}^{-1} &= \frac{1}{2}\int_{\mathrm{\scriptscriptstyle BZ}} 
   \frac{d^{2}k\; dk_{z}}{(2\pi)^{3}} G_{\alpha \alpha}(\vec{k},k_{z}).
\end{align}
The integration of the reciprocal-space elastic Green's
function\cite{Brandt1977a,Brandt1977b,Larkin1979}
\begin{align}\label{eq:G-def}
   G_{\alpha \beta}&(\vec{k},k_{z}) =\\
   &\frac{k_{\alpha}k_{\beta}/k^{2}} {c_{11}(\vec{k},k_{z}) k^{2} 
   + c_{44}(\vec{k},k_{z}) k_{z}^{2}} 
   +\nonumber\frac{\delta_{\alpha \beta} - k_{\alpha}k_{\beta}/k^{2}}
   {c_{66} k^{2} + c_{44}(\vec{k},k_{z}) k_{z}^{2}}
\end{align}
over the vortex lattice Brillouin zone (BZ) involves the dispersive
compression- [$c_{11}(\vec{k},k_{z})$] and tilt-
[$c_{44}(\vec{k},k_{z})$] as well as the non-dispersive shear moduli
($c_{66}$), see also Ref.~\sbonlinecite{Blatter1994}. Here $k = |\vec{k}|$
is the norm of the in-plane momentum $\vec{k}
= (k_{x},k_{y})$. Performing the integration in \eq \eqref{eq:Cbar-def} using
\eq \eqref{eq:G-def} provides us with the expression 
\begin{align}\label{eq:Cbar}
  \C = \nu \frac{a_{0}^{2}}{\lambda}\, \sqrt{c_{66} c_{44}(\vec{0},0)}.
\end{align}
The determination of the numerical factor $\nu$ requires an accurate
evaluation of the linear response of a vortex to a local force. Simple
estimates for $\nu$ can be obtained through the approximate evaluation of the
integral in \eq \eqref{eq:Cbar-def} or by calculating the deformation energy
of a single flux line embedded in a rigid cage potential \cite{Blatter2008}.
In the first case, we neglect the compression term in \eq \eqref{eq:G-def}
since $c_{66} \ll c_{11}$. Using $c_{44}(\vec{k},k_{z}) = c_{44}(\vec{0},0)/[1
+ \lambda^{2}(k^{2} + k_{z}^{2})]$, see Refs.\ \sbonlinecite{Brandt1977b} and
\sbonlinecite{Blatter1994}, and assuming $k^{2}\lambda^{2} \gg
k_{z}^{2}\lambda^{2}$ and $k^{2}\lambda^{2} \gg 1$, we can extend the integral
over $k_z$ to infinity and limit the planar integral to the circularized
Brillouin zone $k^2 < 4\pi/a_0^2$. Using these approximations, we arrive at a
numerical $\nu = 4$.  The alternative estimate is based on a flux line with
elasticity $\varepsilon_l = \varepsilon_0$ trapped within a cage potential
$V_\mathrm{cage} = \pi \varepsilon_0 (u/a_0)^2$ set up by the neighboring
vortices \cite{Blatter2008}, where $\varepsilon_0 =(\Phi_0/4\pi\lambda)^2$
denotes the vortex line energy. Minimizing the total energy $\int dz
[\varepsilon_l u^2/2 + V_\mathrm{cage}(u)] \equiv \bar{C} u^2/2$, we obtain
$\bar{C} = 4 \sqrt{2\pi}\varepsilon_0/a_0$, which corresponds to a factor $\nu
= 4 \sqrt{2}$ when recast into the form \eqref{eq:Cbar}.

Making use of the effective elasticity $\C$, \eq
\eqref{eq:self-consistency-eq-4} can be written in the form
\begin{align}\label{eq:Labusch-equation}
   \bar{C} \vec{u}(\vec{r}) = \vec{f}_{\!p}[\vec{r} + \vec{u}(\vec{r})],
\end{align}
where we have dropped the subscript in the vortex--pin distance, $\vec{r}_{p}
\! \to\! \vec{r}$. It is the appearance of multiple solutions of this
non-linear self-consistency equation which is at the origin of the strong
pinning phenomenon. Inserting the solution $\vec{u}(\vec{r})$ of \eq
\eqref{eq:Labusch-equation} equation back into \eq
\eqref{eq:self-consistency-eq-3}, the displacement field $u_{\alpha}
(\vec{r}_{\nu},z_{\nu})$ of all vortices can be determined.

The self-consistency equation \eqref{eq:Labusch-equation} is easily derived as
the minimizer of the total free energy including contributions from elasticity
and pinning \cite{Koopmann2004},
\begin{align}\label{eq:etot}
   \epin(\vec{r}) &= \frac{1}{2} \bar{C} u^{2} + e_{p}(\vec{r}+\vec{u}).
\end{align}
Indeed, minimizing \eq \eqref{eq:etot} with respect to the displacement $\vec{u}$
leads to \eq \eqref{eq:Labusch-equation}. On the other hand, the derivative
with respect to $\vec{r}$ produces the effective force profile
\begin{align}\label{eq:fne}
   \vec{f}_{\!\mathrm{pin}}(\vec{r}) \equiv - \nabla_{\!\vec{r}}\, \epin(\vec{r})
\end{align}
associated with the total energy \eqref{eq:etot}: evaluating the total
derivative $\nabla_{\!\vec{r}}\, \epin(\vec{r})$, we can express the gradient
$-\nabla_{\!\vec{r}}\, e_{p}(\vec{r})$ through the bare pinning force,
$-\nabla_{\!\vec{r}}\, e_{p}(\vec{r}) = \vec{f}_{\!p}(\vec{r})$, and making
use of \eq \eqref{eq:Labusch-equation}, we find that
\begin{align}\label{eq:fpin}
   \vec{f}_{\!\mathrm{pin}}(\vec{r}) 
   = \vec{f}_{\!p}[\vec{r} + \vec{u}(\vec{r})] = \C\vec{u}(\vec{r}).
\end{align}
Hence, a multi-valued solution of \eq \eqref{eq:Labusch-equation} at a given
$\vec{r}$ entails multi-valued solutions for the energy profile
$\epin(\vec{r})$ as well as the force profile
$\vec{f}_{\!\mathrm{pin}}(\vec{r})$.
\begin{figure}[tb] \includegraphics[width=8.0cm]{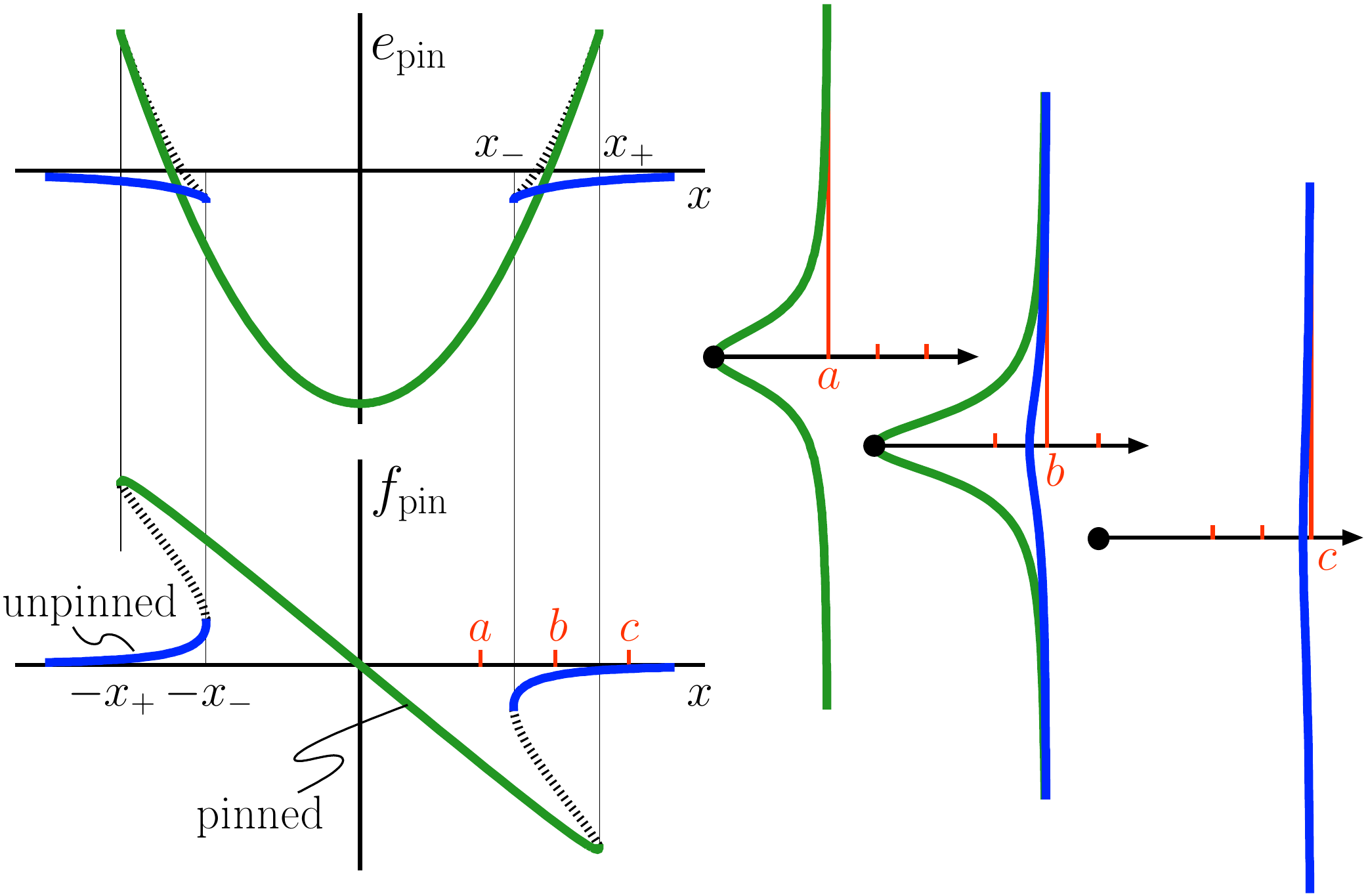}
\caption{Left: Energy- and force profiles at strong pinning $\kappa > 1$ with
multi-valued solutions within the intervals $[-x_{+},-x_{-}]$ and
$[x_{-},x_{+}]$.  Stable pinned (green) and unpinned (blue) branches are
connected by unstable solutions (dashed).  Right: Vortex shapes associated
with pinned ($a,b$ in green) and unpinned ($b,c$ in blue) branches for
different positions away from the pin.}
\label{fig:multivalued}
\end{figure}

For the geometry introduced in \sect \ref{sec:macroscopic-ac-response} and
discussed below, we simplify the formalism further by assuming that all vortex
trajectories with impact parameter $2|y| < t_{\perp}$ experience maximal
pinning, i.e., that of a vortex hitting the defect head-on with $y = 0$. For a
small pinning center, the transverse length $t_\perp$ is of the order of the
vortex core size $\xi$ (the coherence length), while $t_\perp$ is determined
by the pin size for a large defect. With this simplification, the problem
reduces to one effective dimension with \eqs \eqref{eq:Labusch-equation} and
\eqref{eq:fne} taking the form
\begin{align}\label{eq:Labusch-equation-1d}
   \bar{C} u(x) &= f_{p}[x + u(x)] = \fpin(x), \quad \mathrm{with} \\
   \label{eq:fpin-1d}
   \fpin(x) &\equiv -d\epin(x)/dx.
\end{align}

The self-consistency equation \eqref{eq:Labusch-equation-1d} can be easily
tested for multi-valued solutions; these appear when the derivative $du/dx
\equiv u'(x)$ turns infinite. Taking the total derivative of \eq
\eqref{eq:Labusch-equation-1d} with respect to $x$, we find that
\begin{align}\label{eq:uprime}
   u'(x) = \frac{1}{\C/f_{p}'[x+u(x)] - 1}
\end{align}
diverges with increasing pinning force for the first time when the maximal
force derivative $f_{p}'$ matches the elasticity $\bar{C}$. This onset of
strong pinning then is defined by the Labusch criterion $\kappa = 1$, where
$\kappa \equiv \max_{x}[f_{p}'(x)/\C]$ denotes the Labusch parameter. At
small values of $\kappa < 1$, i.e., when $f_{p}'(x) < \C$ for any $x$, the
force profile $\fpin(x)$ is single valued and $u'(x)$ always has the same sign
as $f_{p}'[x + u(x)]$. Within the strong pinning framework, this weak pinning
regime $\kappa < 1$ is associated with a vanishing average pinning force
$F_{\mathrm{pin}} = 0$ and hence $j_{c} = 0$. At the same time, the absence of
force jumps is associated with a divergent Campbell length
$\lambda_{\mathrm{\scriptscriptstyle C}} = \infty$, see below. These results
are modified if collective phenomena are included in the model, a topic that
goes beyond the present work.

For $\kappa > 1$, the pinning force profile $\fpin(x)$ turns multivalued with
inflection points at $\pm x_{-}$ and $\pm x_{+}$ ($0 < x_{-} < x_{+}$) where
\eq \eqref{eq:uprime} diverges, see \fig
\ref{fig:multivalued}. Between the two inflection points $x_{\pm}$, two stable
and one unstable branch exist, the latter being characterized by
$u'(x)/f_{p}'[x+u(x)] < 0$. The two stable branches are smoothly connected to
a vortex trapped by the pin and a vortex detached from the pin, respectively.
Hence, we shall use the terminology `pinned' and `unpinned' branches for these
two solutions. The pinning region $|\vec{r}| < x_{-}$, where only the pinned
branch exists, defines the transverse trapping length $t_{\perp} = 2 x_{-}$.
For strong pinning, the pinscape produces a finite macroscopic pinning force
density by asymmetrically populating the different branches of $\fpin(x)$.

Hence, the Labusch criterion \cite{Labusch1969} $\kappa = f_{p}'(x_{m}) /\C$ $
= 1$, with $\max_{x}[f_{p}'(x)] = f_{p}'(x_{m})$ realized at $x_{m}$, serves
as a quantitative separation between the regimes of weak ($\kappa < 1$) and
strong ($\kappa > 1$) pinning. As pinning vanishes alltogether at $\kappa <
1$, one often uses the distinction between weak, intermediate, and strong
`strong-pinning' regimes with different $j_c$-scalings, $j_c \propto (\kappa-1)^2$
and $j_c \propto \kappa^2$, at the two extremes. With the formalism of strong
pinning at hand, we are now ready to discuss the physical implications of
vortex pinning by a low density of pinning centers.

\subsection{Critical current}\label{sec:critical-current}
For randomly (and homogeneously) distributed pinning sites with a small
density $n_p$ (see below for a quantitative criterion), the \emph{macroscopic}
pinning force density $\Fpin$ results from proper averaging of the
\emph{microscopic} forces \eqref{eq:fpin-1d},
\begin{align}\label{eq:Fpin}
   \Fpin = n_{p}\av{\fpin} = n_{p}\frac{t_{\perp}}{a_{0}} \int \frac{dx}{a_{0}} 
   \fpin(x)\big|_{\mathrm{o}}
\end{align}
with $\fpin(x)|_{\mathrm{o}} \equiv f_{p}[x + u^{\mathrm{o}}(x)]$ referring to
the occupied branches $u^{\mathrm{o}}$ in the effective force profile
$\fpin(x)$. In the zero-field-cooled (critical) state, the pinning landscape
acts with the critical force density $\Fpin = -F_{c}$ against the Lorentz
force density $j B_0/c$, thus defining the critical current density $j_{c} = c
F_{c}/B_{0}$.  This maximal pinning force density is achieved when the pinned
branch $u^{\mathrm{p}}(x)$ is occupied in a maximally asymmetric way between
$-x_{-}$ to $x_{+}$, see \fig \ref{fig:occupation-shift}. Combining \eqs
\eqref{eq:fpin-1d} and \eqref{eq:Fpin}, we arrive at a microscopic expression
for the critical current $j_{c} = -(c/B_{0}) F_c$,
\begin{align}\label{eq:jc}
   j_{c} = \frac{c n_{p} t_{\perp}}{\Phi_{0}} \Delta \epin,
\end{align}
where $\Delta \epin$ denotes the sum of the jumps in $\epin(x)$ between pinned
and unpinned branches at the positions $-x_{-}$ and $x_{+}$. More precisely,
\begin{align}\label{eq:depin}
   \Delta \epin = (\epin^{\mathrm{up}} - \epin^{\mathrm{p}})\big|_{-x_-}
                + (\epin^{\mathrm{p}} - \epin^{\mathrm{up}})\big|_{ x_+},
\end{align}
where the superscripts `up' and `p' denote unpinned and pinned branches.
While \eq \eqref{eq:jc} provides a quantitative expression for the critical
current density within strong pinning theory, we may use $t_{\perp} \sim \xi$
and $\Delta \epin \sim f_p x_+ \sim f_{p}^{2}/\C$ (with $f_{p}$ the typical
strength of the bare pinning force and $x_+ \sim \kappa \xi$) to arrive at a
qualitative estimate for the critical current, $j_{c} \sim c n_{p} \xi
f_{p}^{2}/ \Phi_{0} \C$. Together with the scaling $\C \propto
(\Phi_{0}/4\pi\lambda)^2/a_0 = \varepsilon_0/a_0$ of the effective elasticity
at low fields, we arrive at 
\begin{align}\label{eq:jc_est}
   j_{c} \sim j_{\mathrm{dp}}\,  (n_{p} a_{0} \xi^{2})\,  (\kappa\xi/a_0)^{2},
\end{align}
with the depairing current $j_{\mathrm{dp}} = c \Phi_{0} / (12
\sqrt{3} \pi^{2} \lambda^{2}\xi)$ and the small parameter $n_{p} a_{0} \xi^{2}
\ll 1$ defining the regime of 3D strong pinning, see Ref.\
\sbonlinecite{Blatter2004}. The field scaling $j_{c} \propto 1/\sqrt{B_{0}}$
is in agreement with the results obtained in the early work on strong pinning
by Ovchinnikov and Ivlev\cite{Ovchinnikov1991}.

\subsection{Campbell Length}\label{sec:campbell-length}
The Campbell penetration depth $\lambda_{\mathrm{\scriptscriptstyle C}}$ is
another measureable quantity characterizing the pinning landscape.  In a
microscopic derivation of $\lambda_{\mathrm{\scriptscriptstyle C}}$, we have
to find the dynamical change in pinning force $\delta \Fpin[U(X,t)]$.  The
latter is determined by the change in branch occupation due to the macroscopic
displacement $U$ of the vortex lattice. As shown below, the macroscopic
Campbell length relates to $\Delta \fpin$, the sum of the
jumps in $\fpin(x)$ between occupied and unoccupied branches,
\begin{align}\label{eq:Campbell}
   \frac{1}{\lambda_{\mathrm{\scriptscriptstyle C}}^{2}} &=
      \frac{4\pi n_{p} t_{\perp}}{B_{0} \Phi_{0}} \Delta \fpin,
\end{align}
and hence probes another quantity than $j_c$, \eq \eqref{eq:jc}, Beloow, we
derive this result for two initial states of particular importance, the
zero-field-cooled (or Bean critical) state and the field-cooled state.
\subsubsection{Bean critical state}\label{sec:zfc}
\begin{figure}[bt]
\includegraphics[width=7.0cm]{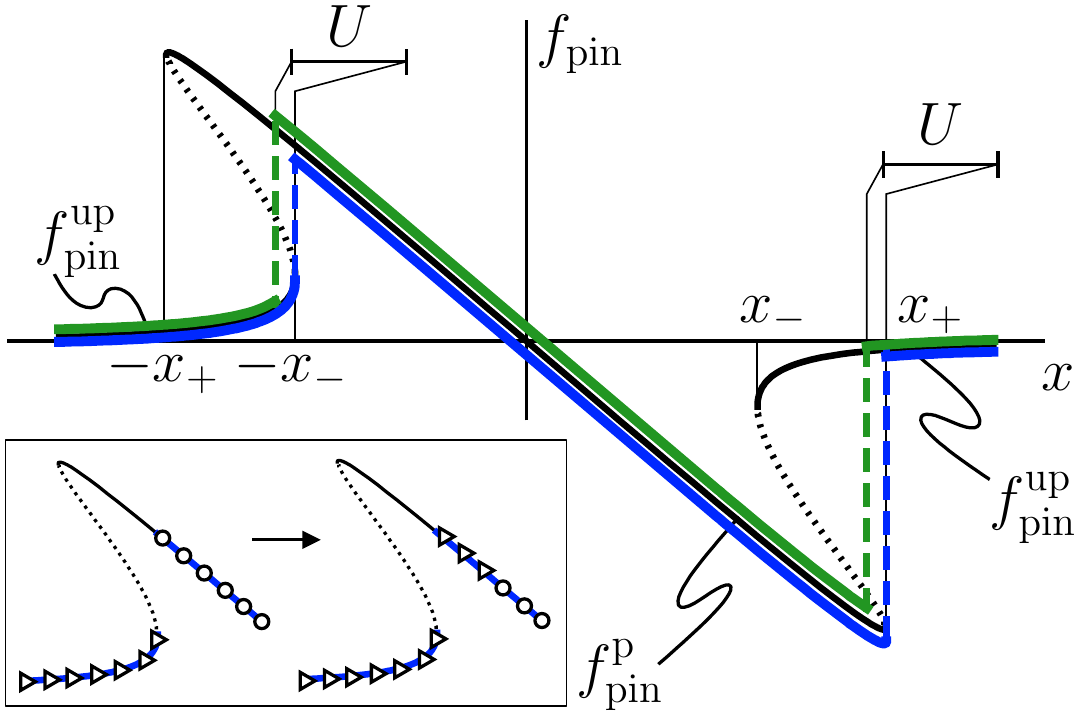}
\caption{Occupation of the pinned ($\fpin^{\mathrm{p}}$) and unpinned
($\fpin^{\mathrm{up}}$) branches in the critical state (blue). A macroscopic
displacement of all vortices to the left, $U < 0$, results in a new branch
occupation (green), with vortices populating the pinned (unpinned) branch
below $-x_{-}$ ($x_{+}$) while the corresponding states on the unpinned
(pinned) branch get depleted; this change in occupation leads to a decrease of
the macroscopic pinning force density. On the other hand, for a uniform shift
of all vortices to the right with $U > 0$, vortices within a distance $U$ to
the left of the branch edge irreversibly jump to the other stable solution
(see inset). This process does not lead to a net change in branch occupation
and the pinning force remains at its critical value.}
\label{fig:occupation-shift}
\end{figure}

The Bean critical state, as described in \sect \ref{sec:critical-current}, is
characterized by the maximal or critical pinning force $\Fpin = -F_{c}$ with
an asymmetric branch occupation. Hence, depending on the sign of $U$, the
branch occupation will be affected differently. Specifically, for a
macroscopic shift of all vortices in the direction of the Lorentz force, i.e.,
$x \to x + U$ with $U > 0$, most vortices adiabatically follow their
branch, $u(x) \to u(x+U)$. The few vortices on the unpinned (pinned) branch
at distances less than $U$ away from the branch edge at $-x_{-}$ ($x_{+}$)
will be pushed beyond that boundary and irreversibly jump to the pinned
(unpinned) solution, see inset of \fig \ref{fig:occupation-shift}. Hence, a
displacement $U > 0$ leads to (i) a net penetration of vortices into the
sample while (ii) leaving the branch occupation unchanged, i.e.,
\begin{align}\label{eq:FofUpos}
   \delta \Fpin(U > 0) = 0.
\end{align}
On the other hand, for a displacement $x \to x + U$ to the left with $U < 0$,
i.e., against the critical slope, all vortices adiabatically follow their
branches. The occupation of the pinned branch then is shifted to lie between
$-x_{-}+U$ and $x_{+}+U$, see \fig \ref{fig:occupation-shift}. Similarly, the
unpinned branch is occupied until $-x_- +U$ and onwards from $x_+ +U$. The
change in pinning force $\delta \Fpin(U)$ is obtained from the difference of
the pinning force \eqref{eq:Fpin} evaluated for the critical state shifted by
$U$, leading to
\begin{align}
   \!\!\!
   \delta \Fpin(U < 0) &= \frac{n_{p}t_{\perp}}{a_{0}^{2}}
                    \bigg[\!\!\!\!\int\limits_{-x_{-}+U}^{-x_{-}}\!\!\!\!\!
                           dx\, [\fpin^{\mathrm{p}}(x) - \fpin^{\mathrm{up}}(x)]\;\;
                           \\[-1em]\nonumber
                    &\qquad\qquad\qquad+\int\limits_{x_{+}+U}^{x_{+}}\!\!\!
                           dx\, [\fpin^{\mathrm{up}}(x) - \fpin^{\mathrm{p}}(x)]
                           \bigg]\\ \nonumber
             &\!\!\!\!\!\!\!\!\!\!\!\!\!\!\!\!\!
               \approx -\frac{n_{p}t_{\perp}}{a_{0}^{2}}
                  \big[(\fpin^{\mathrm{p}} - \fpin^{\mathrm{up}})\big|_{-x_{-}}
                  \! + (\fpin^{\mathrm{up}}- \fpin^{\mathrm{p}})\big|_{x_{+}}\big] U\\
             \label{eq:FofUneg}
             &\!\!\!\!\!\!\!\!\!\!\!\!\!\!\!\!\!
               = -\frac{n_{p}t_{\perp}}{a_{0}^{2}}\Delta \fpin\, U.
\end{align}
This drop in the critical force always appears when vortices start moving to
the left and is associated with a reduction of $U(X,t)$ with increasing time.
In order to follow dynamically the appearance and disappearance of this term
in the equation of motion, we introduce the max-field $U_{0}(X,t) =
\max_{t'<t}U(X,t)$. Whenever $U(X,t)$ changes the direction of motion from
right to left, $U$ starts deviating from $U_{0}$. The argument $U$ in \eq
\eqref{eq:FofUneg} should then be replaced by $U - U_{0}$. In the end, the
piecewise change of the pinning force entering the macroscopic equation of
motion \eqref{eq:eom} reduces to the simple expression
\begin{align}\label{}
   \delta \Fpin(U) = - \alpha_{\mathrm{sp}} (U - U_{0}),
\end{align}
and satisfies both \eqs \eqref{eq:FofUpos} and \eqref{eq:FofUneg}. In the
above expression, the underlying pinning potential enters solely through the
coefficient $\alpha_{\mathrm{sp}} = (n_{p}t_{\perp}/a_{0}^{2})\Delta \fpin$.
This coefficient can be understood as the mean curvature of the pinning
energy,
\begin{align}\label{eq:curv}
   \alpha_{\mathrm{sp}} = n_p \langle e''_\mathrm{pin} \rangle 
   = - \frac{n_p t_\perp}{a_0} \int \frac{dx}{a_0} \, 
   f'_\mathrm{pin}(x)|_{\mathrm{o}},
\end{align}
similar to the mean force in Eq.\ \eqref{eq:Fpin} defining the critical
current density $j_c$.

The asymmetric response of the vortex system to an increasing versus
decreasing field is associated with a peculiar transient initialization
towards a periodic vortex motion, where on every $ac$ cycle vortices are
pumped (and diffusively penetrate) into the sample, asymptotically changing
the $dc$ field from $B_{0}$ to $B_{0} + h_{ac}$ after many cycles. A detailed
discussion of this process is presented in Ref.~\sbonlinecite{Willa2015b},
where it is shown that the number of cycles needed to shift the critical state
from $B_{0}$ to $B_{0} + h_{ac}$ within the depth of the Bean profile $L = c
B_{0}/4\pi j_{c}$ is about $N = (\pi L / 2\ell_{D})^2$, with $\ell_{D} =
(B_0^{2}/2 \omega\eta)^{1/2}$ the diffusion length per cycle period
$2\pi/\omega$. After this rectification process, vortices move reversibly in
their wells as $U$ always remains below $U_{0}$. The latter reaches the
asymptotic form $U_{0}(X) = (h_{ac}/B_{0})(L - X)$. For a sample of finite
thickness $d < 2L$ along $X$, $d/2$ replaces $L$ is the expressions for $N$
and $U_{0}$. The reversible dynamics of vortices after the initialization can
be solved by substituting the variable $\delta U(X,t) \equiv U(X,t) -
U_{0}(X)$ into \eq \eqref{eq:eom} and one finds
\begin{align}\label{}
   \delta U(X,t) = - \lambda_{\mathrm{\scriptscriptstyle C}} (h_{ac}/B_{0})
                     e^{-X/\lambda_{\mathrm{\scriptscriptstyle C}}}
                     \big[ 1 - e^{- i \omega t}\big].
\end{align}
The $ac$ response of the vortex lattice in the critical state
\begin{align}\label{}
   \delta B(X,t) = h_{ac} e^{-X/\lambda_{\mathrm{\scriptscriptstyle C}}}
                     e^{- i \omega t}
                 + h_{ac}[1 - e^{-X/\lambda_{\mathrm{\scriptscriptstyle C}}}]
\end{align}
is regular and involves the Campbell length
$\lambda_{\mathrm{\scriptscriptstyle C}}$ given by \eq \eqref{eq:Campbell}.
The asymptotic solution consists of an oscillatory response within a surface
layer $\sim \lambda_{\mathrm{\scriptscriptstyle C}}$ and a rectified $dc$ part
that has penetrated deep into the bulk. This behavior is very similar to the
critical $ac$ response discussed by Bean \cite{Bean1962} (Bean penetration),
where large amplitude oscillations $h_{ac} \gg j_{c} \lambda_{\mathrm{
\scriptscriptstyle C}} / c$ generate a nonlinear response. An extended
comparison between these two scenarios is given in Ref.\
\sbonlinecite{Willa2015b}.

\subsubsection{Field-cooled state}\label{sec:fc}
The field-cooled state is characterized by vanishing net currents and net
pinning forces. In the strong pinning regime, the vanishing pinning force
translates into a symmetric occupation of the branches, with jumps between the
pinned and unpinned branch located at $\pm x_{\mathrm{jp}} \in [x_{-},
x_{+}]$. If this position is away from the branch edges $x_{\pm}$, the
oscillation is always reversible. If $x_{\mathrm{jp}}$ coincides with one of
the branch edges $x_{\pm}$, a one-cycle initialization process reshuffles few
vortices near the branch edges, after which the oscillation is reversible and
the result in \eq \eqref{eq:Campbell} involves the jumps at $\pm
x_{\pm}$. Hence, no complex initialization process needs to be studied for
the field-cooled situation and the change in the pinning force is always given
by the expression
\begin{align}\label{eq:delta-Fpin-FC}
   \delta \Fpin(U) = -\frac{n_{p}t_{\perp}}{a_{0}^{2}} \Delta \fpin U,
\end{align}
with $\Delta \fpin$ now involving two identical jumps at $\pm
x_{\mathrm{jp}}$.

In order to quantify the Campbell length in the field-cooled state, the
central remaining task is to determine the precise position of the jump
$x_{\mathrm{jp}}$ within the interval $[x_{-}, x_{+}]$ and find the
corresponding jump $\Delta \fpin$ upon changing the temperature $T$. For
insulating or metallic inclusions, where pinning smoothly increases when
crossing $H_{c2}(T)$ (see below), we can follow the branch occupation as a
function of $T$ as the system evolves from weak $(\kappa <  1)$ to strong
$(\kappa > 1)$ pinning through the Labusch point
$\kappa(T_{\mathrm{\scriptscriptstyle L}}) = f_{p}'(x_{m})/\C
|_{T_{\mathrm{\scriptscriptstyle L}}} =  1$ defining the Labusch temperature
$T_{\mathrm{\scriptscriptstyle L}}$ (here, $x_m$ denotes the point of maximal
slope $f_{p}'$). Above the Labusch temperature, $T >
T_{\mathrm{\scriptscriptstyle L}}$, the force profile is single valued, the
critical current vanishes [see \eq \eqref{eq:jc}], and the Campbell length is
formally infinite [see \eq \eqref{eq:Campbell}] due to the absence of jumps.
These singular results are regularized once collective pinning effects are
considered. Upon lowering the temperature, the system reaches the Labusch
point $\kappa(T_{\mathrm{\scriptscriptstyle L}}) = 1$, where the pinning force
$\fpin$ develops a vertical slope at $x_{0\mathrm{\scriptscriptstyle L}}$,
$f_p'[x_{0\mathrm{\scriptscriptstyle L}} + u(x_{0\mathrm{\scriptscriptstyle
L}})] = \C$, see \eq \eqref{eq:uprime}. The combination with \eq
\eqref{eq:Labusch-equation-1d} and the Labusch criterion $f_{p}'(x_{m}) = \C$
provides us with the relation $x_{0\mathrm{\scriptscriptstyle L}} = x_{m} -
f_{p}(x_{m})/\C |_{T_{\mathrm{\scriptscriptstyle L}}}$. Lowering the
temperature $T$ further below $T_{\mathrm{\scriptscriptstyle L}}$, the pinning
force $\fpin$ turns multi-valued within the intervals $\pm [x_{-}, x_{+}]$,
see \fig \ref{fig:multivalued}.

\begin{figure}[tb]
\includegraphics[width=5.0cm]{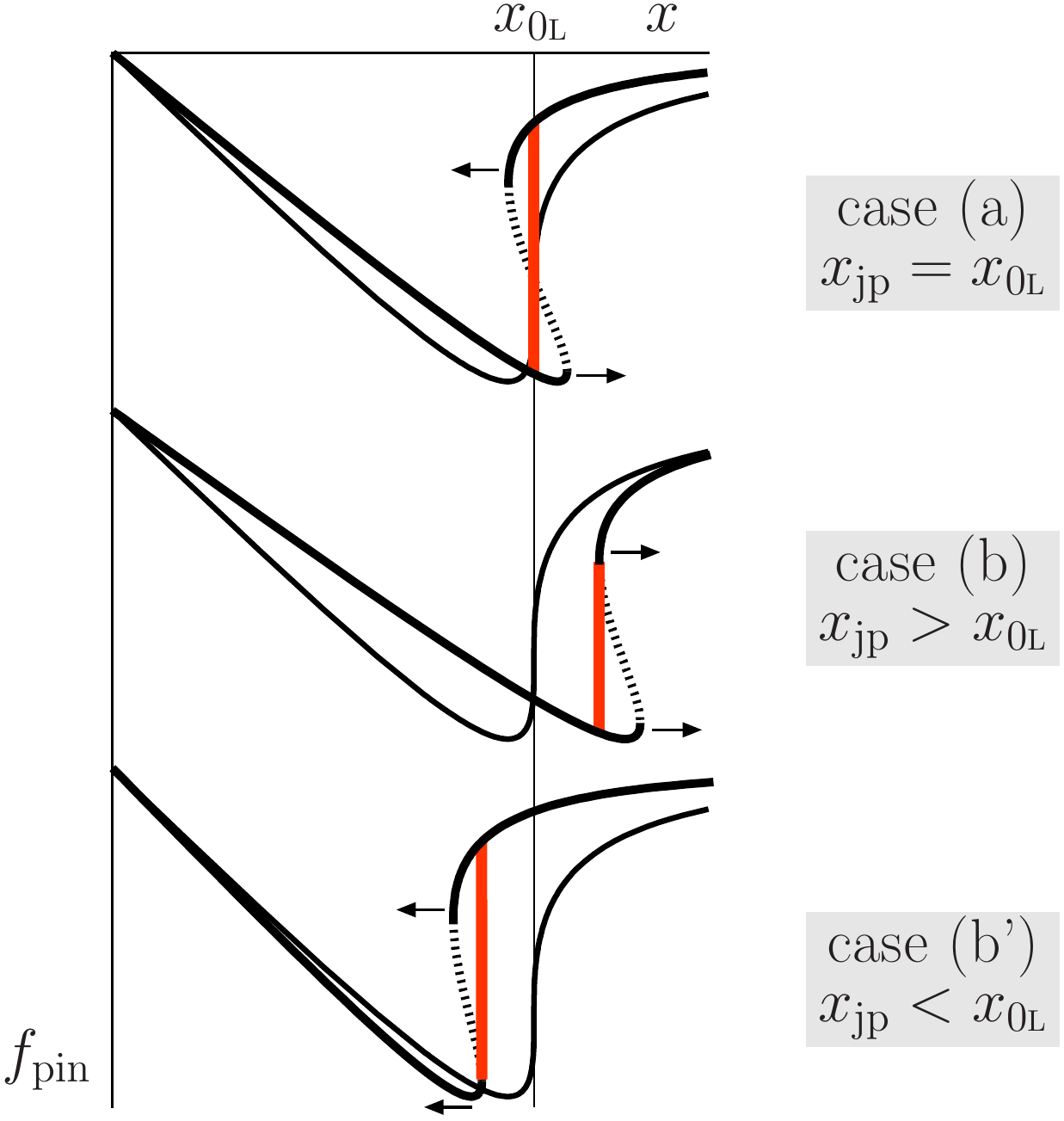}
\caption{Upon lowering the temperature below the Labusch temperature, $T <
T_{\mathrm{\scriptscriptstyle L}}$, the force profile turns multi-valued with
the bistable region centered about $x_{0\mathrm{\scriptscriptstyle L}}$ [case
(a)], to its right [case (b)], or to its left [case (b')]. Specific
microscopic pinning mechanisms (see text) entail one of these three cases,
each associated with its force jump at $x_{\mathrm{jp}}$. These force jumps
$\Delta f_\mathrm{pin}$ (red lines) are probed by a measurement of the
Campbell length $\lambda_{\mathrm{ \scriptscriptstyle C}}$.}
\label{fig:cases-a-b-c}
\end{figure}

Depending on the temperature dependence of the elastic and pinning forces, we
have identified three possible scenarios defining the (symmetric) jump
positions $\pm x_{\mathrm{jp}}$ in the branch occupation, see Fig.\
\ref{fig:cases-a-b-c}. In the first case (a), the branch edges $x_{\pm}$ move
away from $x_{0\mathrm{\scriptscriptstyle L}}$ in opposite directions, $x_{-}
< x_{0\mathrm{\scriptscriptstyle L}} < x_{+}$. The second case (b), describes
the situation where both boundaries $x_{\pm}$ become larger than
$x_{0\mathrm{\scriptscriptstyle L}}$ upon cooling,
$x_{0\mathrm{\scriptscriptstyle L}} < x_{-} < x_{+}$, while they become
smaller in the third case (b'), $x_{-} < x_{+} <
x_{0\mathrm{\scriptscriptstyle L}}$.

In the simplest case (a), vortices at $x < x_{0\mathrm{\scriptscriptstyle L}}$
($x > x_{0\mathrm{\scriptscriptstyle L}}$) follow adiabatically the evolution
of the pinned (unpinned) branch and the occupation jumps at $x_{\mathrm{jp}} =
x_{0\mathrm{\scriptscriptstyle L}}$, such that $\Delta \fpin = 2\Delta
\fpin|_{x_{0\mathrm{\scriptscriptstyle L}}}$ enters the expression
\eqref{eq:Campbell} for the Campbell length, with $\Delta \fpin|_{x_{0\mathrm{
\scriptscriptstyle L}}}$ denoting the jump in $\fpin$ at
$x_{0\mathrm{\scriptscriptstyle L}}$.
In case (b), the unpinned branch, initially existing for $x >
x_{0\mathrm{\scriptscriptstyle L}}$, becomes unstable in the interval
$[x_{0\mathrm{\scriptscriptstyle L}}, x_{-}]$. As a result, vortices with
$x_{0\mathrm{\scriptscriptstyle L}} < x < x_{-}$ now occupy the pinned branch
and the force jump appears at $x_{\mathrm{jp}} = x_{-}$, with $\Delta \fpin =
2\Delta \fpin|_{x_{-}}$ entering $\lambda_{\mathrm{\scriptscriptstyle C}}$.
Similarly, in case (b'), the vortices populate the unpinned branch in the
interval $[x_{+},x_{0\mathrm{\scriptscriptstyle L}}]$ where the pinned
solution has stopped existing. The Campbell length then involves the jumps at
$\pm x_{+}$.

\begin{figure}[b!]
\includegraphics[width=7.0cm]{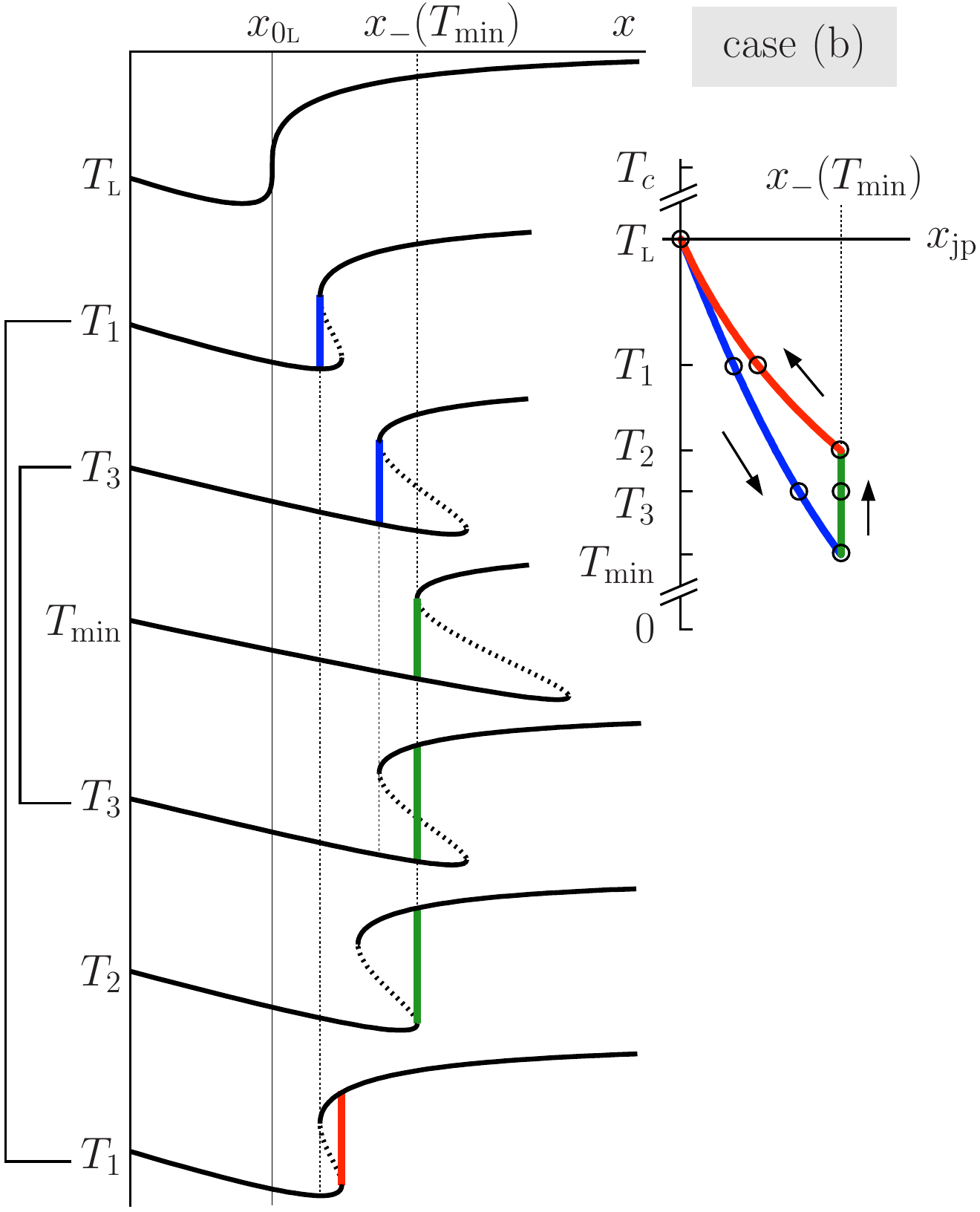}
\caption{Illustration of the evolution of the force profile at different
temperatures $T_{\mathrm{\scriptscriptstyle L}} > T_{1} > T_{2} > T_{3} >
T_{\mathrm{min}}$ [case (b) in Fig.\ \ref{fig:cases-a-b-c}], and the
associated hysteresis in the force jump position $x_{\mathrm{jp}}$ (inset)
upon cooling from $T_{\mathrm{\scriptscriptstyle L}}$ down to
$T_{\mathrm{min}}$ and subsequent heating. Upon cooling, the force jump probed
by the Campbell length is positioned at the branch edge $x_{\mathrm{jp}} =
x_{-}$ (blue). During reheating, the jump's position first remains fixed at
$x_{\mathrm{jp}} = x_{-}(T_{\mathrm{min}})$ (green). At the temperature $T_2$,
the edge at $x_+$ coincides with the position $x_-$ previously reached at
$T_\mathrm{min}$. When the temperature increases above $T_{2}$ the force
jump follows the other branch edge at $x_{+}$ (red).}
\label{fig:hysteresis}
\end{figure}

The repopulation of vortices from the unpinned to the pinned branch in case
(b) [or vice-versa in case (b')] leads to a hysteretic response if the system
is reheated after the cooling process. Consider a system in case (b) cooled
to the minimal temperature $T_{\mathrm{min}}$ and subsequently reheated. Upon
cooling, vortices on the unpinned branch become locally unstable at $x_{-}(T)$
and the jump in occupation follows $x_{-}(T)$; the Campbell length \eq
\eqref{eq:Campbell} involves $\Delta \fpin = 2 \Delta \fpin |_{x_{-} (T)}$,
see the discussion above and Fig.\ \ref{fig:hysteresis}. Upon reversing the
temperature sweep at $T_\mathrm{min}$, the jump is locked to $x_{-}
(T_{\mathrm{min}})$ as all vortices remain stable within their branches; the
Campbell length now involves the jumps $\Delta \fpin = 2\Delta \fpin
|_{x_{-}(T_\mathrm{min})}$. With the temperature increasing further, vortices
on the pinned branch become locally unstable at $x_{+}(T) \leq
x_{-}(T_{\mathrm{min}})$ and the jump in occupation follows $x_{+}(T)$; the
Campbell length then involves the jumps $\Delta \fpin = 2\Delta \fpin
|_{x_{+}(T)}$. The difference in the force jumps then naturally leads to a
hysteretic behavior of the Campbell length
$\lambda_{\mathrm{\scriptscriptstyle C}}$.

In order to provide a quantitative insight into the evolution of $x_{\pm}$
away from (but close to) $x_{0\mathrm{\scriptscriptstyle L}}$, we expand the
bare pinning force around $x_{m}$,
\begin{align}\label{eq:expansion}
   f_{p}(x) \approx f_{p}(x_{m})
                    + f_{p}'(x_{m})  (x - x_{m})
                    - \frac{\gamma}{3}(x - x_{m})^{3},
\end{align}
with $f_{p}''(x_{m}) = 0$ and $\gamma \equiv -f_{p}'''(x_{m})/2 > 0$. These
two conditions originate from the definition of $x_{m}$ as the location
maximizing $f_{p}'(x)$. Using the above expression in combination with \eqs
\eqref{eq:Labusch-equation-1d} and \eqref{eq:fpin-1d}, allows us to find the
inflection points $x_{\pm}$ in $u(x)$ [or $\fpin(x)$] characterized through a
vanishing denominator on the right-hand side of \eq \eqref{eq:uprime}. A
straightforward calculation provides the result
\begin{align}\label{eq:xpm}
   x_{\pm} = x_{0} \pm \frac{2}{3}\sqrt{\frac{\C}{\gamma}}(\kappa - 1)^{3/2}
\end{align}
where $x_{0}(T) = x_{m} - f_{p}(x_{m})/\C |_{T}$ is a generalization of
$x_{0\mathrm{\scriptscriptstyle L}} = x_{0}(T_{\mathrm{\scriptscriptstyle
L}})$. Case (a) is realized when $x_{0}(T) = x_{0\mathrm{\scriptscriptstyle
L}}$ to order $(\kappa -1)^2$, i.e., the relevant jump is at
$x_{0\mathrm{\scriptscriptstyle L}}$. Solving the self-consistency equation
\eqref{eq:Labusch-equation-1d} for $u_\pm(x_{0\mathrm{\scriptscriptstyle L}})$
and using the relation $\Delta u|_{x_{0\mathrm{\scriptscriptstyle L}}} \equiv
[u_+ - u_-]_{x_{0\mathrm{ \scriptscriptstyle L}}} = \Delta
\fpin|_{x_{0\mathrm{ \scriptscriptstyle L}}} / \C$, we find the force jump
\begin{align}\label{eq:fpin_x0L}
   \Delta \fpin|_{x_{0\mathrm{\scriptscriptstyle L}}} 
   = 2\sqrt{3} \C \sqrt{{\C}/{\gamma}} (\kappa - 1)^{1/2}.
\end{align}

However, case (a) is a special situation since the term linear in $(\kappa-1)$
in the expansion of $x_{0}$, see \eq \eqref{eq:xpm},
has to vanish. For the generic cases (b) and (b') the relevant jumps are at
$x_{-}$ and $x_{+}$, respectively. Solving the cubic self-consistency
equation \eq \eqref{eq:Labusch-equation-1d} for $u(x_{+})$, we find one doubly
degenerate solution $u_{+}$ for the deformation at the edge of the pinned
branch and a non-degenerate solution $u_{-}$ within the unpinned branch. The
force jump at $x_{+}$ then amounts to $\C |u_{+} - u_{-}|$. A similar
analysis can be carried out for $u(x_{-})$; to this order in the expansion
\eqref{eq:expansion}, both jumps turn out identical and take the form
\begin{align}\label{eq:fpin_xpm}
   \Delta \fpin|_{x_{\pm}} = 3\C \sqrt{{\C}/{\gamma}} (\kappa - 1)^{1/2}.
\end{align}

Within the same approximation, the Campbell length in the Bean critical state
involves one jump at $-x_{-}$ and one at $x_{+}$; since both jumps are equal,
the result for the zero-field-cooled state coincides with that for the
field-cooled states of cases (b) and (b'). On the other hand, the non-generic
case (a) features a smaller Campbell length since the associated jump centered
between $x_{-}$ and $x_{+}$ is larger by a factor $2/\sqrt{3} \approx 1.155$.

Away from the Labusch point, the degeneracy of the cases (b) and (b') is
removed and all force jumps are different, with
$\lambda_{\mathrm{\scriptscriptstyle C}}|_{\mathrm{\scriptscriptstyle FC(b')}} <
\lambda_{\mathrm{\scriptscriptstyle C}}|_{\mathrm{\scriptscriptstyle ZFC}} <
\lambda_{\mathrm{\scriptscriptstyle C}}|_{\mathrm{\scriptscriptstyle FC(b)}}$. Simple
expressions can be provided in the limit $\kappa \gg 1$, where the force jump
at $x_{\mathrm{jp}}$ assumes the approximate value $\Delta
f_\mathrm{pin}|_{x_\mathrm{jp}} \simeq \C x_{\mathrm{jp}}$ (note that the
pinned branch is well described by $f_\mathrm{pin} = -\C x$ away from the
Labusch point). These jumps appear at $x_{+} = \kappa \xi$ for case (b') and
$x_{0\mathrm{\scriptscriptstyle L}} \simeq \xi$ for case (a). The jump at
$x_{-}(\kappa)$ for case (b) depends on the tail of the force profile
$f_{p}(x)$ far from the pin. For a power-law decay $f_{p}(x) \propto
-(\xi/x)^{n}$, we find $x_{-}(\kappa) \simeq \xi \kappa^{1/(n+1)}$, while for
an exponential tail $f_{p}(x) \propto - \exp(-x/\xi)$, we have $x_{-}(\kappa)
\simeq \xi \ln(\kappa)$. We then find
\begin{align}\label{eq:lambda_ratios}
   \frac{\lambda_{\mathrm{\scriptscriptstyle C}}^2|^{}_{\mathrm{\scriptscriptstyle FC(b')}}}
     {\lambda_{\mathrm{\scriptscriptstyle C}}^2|^{}_{\mathrm{\scriptscriptstyle ZFC}}}
        \! \simeq \! \frac{1}{2}, ~~~
   \frac{\lambda_{\mathrm{\scriptscriptstyle C}}^2|^{}_{\mathrm{\scriptscriptstyle FC(b)}}}
     {\lambda_{\mathrm{\scriptscriptstyle C}}^2|^{}_{\mathrm{\scriptscriptstyle ZFC}}}
        \! \simeq \! \frac{\kappa\xi}{2x_{-}(\kappa)}, ~~~
   \frac{\lambda_{\mathrm{\scriptscriptstyle C}}^2|^{}_{\mathrm{\scriptscriptstyle FC(a)}}}
     {\lambda_{\mathrm{\scriptscriptstyle C}}^2|^{}_{\mathrm{\scriptscriptstyle ZFC}}}
        \! \simeq \! \frac{\kappa}{2},
\end{align}
with 
\begin{align}\label{eq:lambda_scale}
  \frac{1}{\lambda_{\mathrm{\scriptscriptstyle C}}^{2}|^{}_{\mathrm{\scriptscriptstyle ZFC}}} 
   \simeq \frac{4\pi n_{p} t_{\perp}}{B_{0} \Phi_{0}} \C \kappa \xi.
\end{align}
If one neglects the weak dependence of $t_\perp = 2 x_-$ on $\kappa$, the two
last relations in \eqref{eq:lambda_ratios} tell that the information on the
pinning force $f_p$ disappears from the Campbell length in the field-cooled
cases (a) and (b). More precisely, following the discussion above, the
residual weak dependence of $x_-$ on $\kappa$ provides information on the
decay of $f_p(x)$ at large $x > \xi$.

Applying the same estimates as in \sect \ref{sec:critical-current}
to the zero-field-cooled Campbell length in \eq \eqref{eq:lambda_scale},
we arrive at the qualitative scaling 
\begin{align}\label{eq:lambda_est}
   \lambda_{\mathrm{\scriptscriptstyle C}}^{2}|^{}_{\mathrm{\scriptscriptstyle ZFC}}  \sim
   \frac{\lambda^{2}}{n_{p} a_{0} \xi^{2} \, \kappa} > \lambda^{2}.
\end{align}
Similar to the critical current, see \sect \ref{sec:critical-current}, the
Campbell length involves the small parameter $n_{p} a_{0} \xi^{2} \ll 1$
characteristic of the 3D strong pinning limit\cite{Blatter2004}.

\subsubsection{High fields}\label{sec:high-fields}

\begin{figure}[tb]
\includegraphics[width=6.0cm]{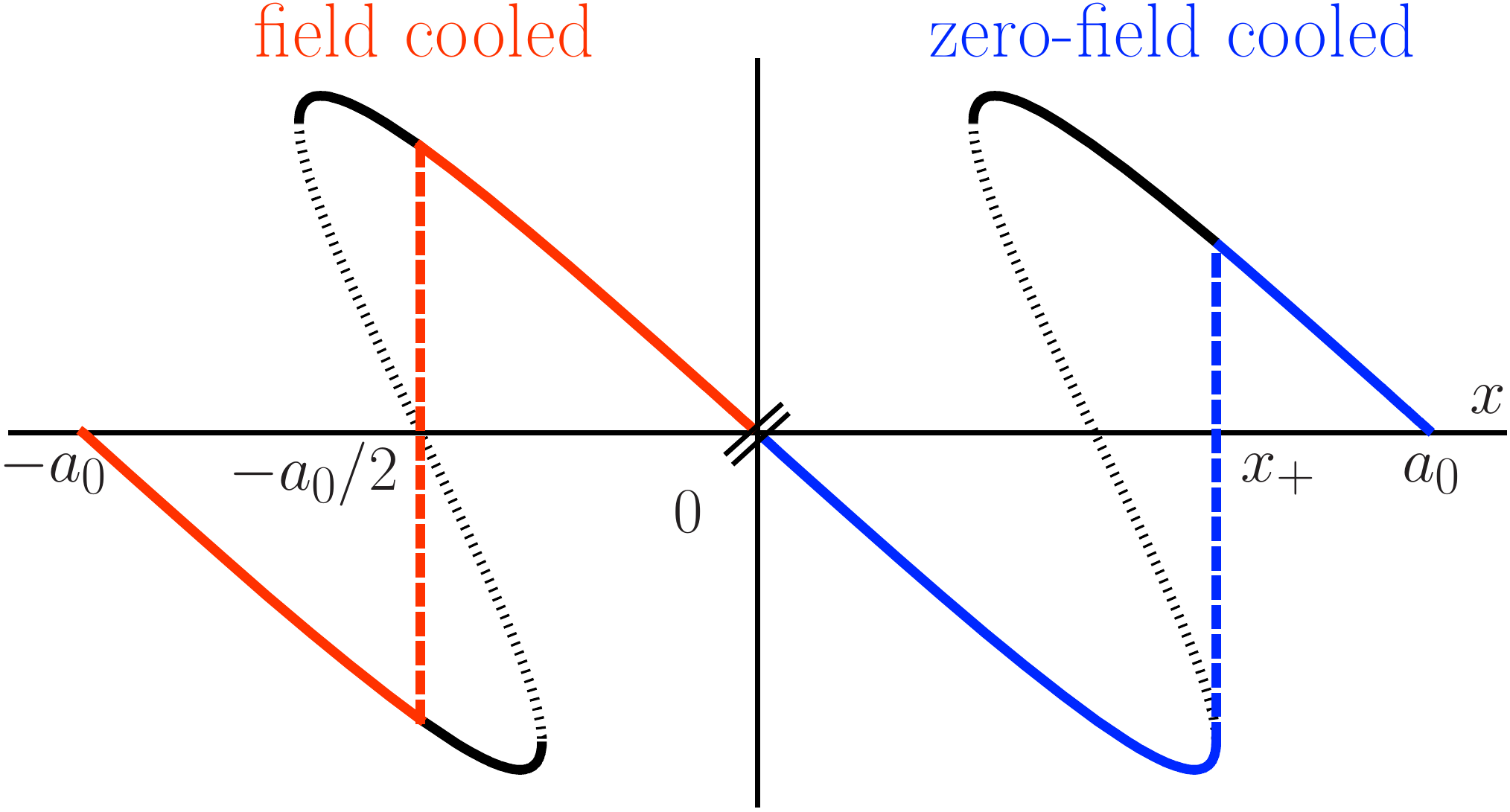}
\caption{Branch occupation at high fields when the vortex system is
prepared in the field-cooled (left) and zero-field-cooled (right) states. At
high fields, the unpinned branch has disappeared and the Campbell length
involves a single jump in force $\Delta \fpin = \Delta \fpin
|_{x_{\mathrm{jp}}}$ at $x_{\mathrm{jp}} = -a_{0} / 2$ ($x_{\mathrm{jp}} =
x_{+}$) indicated with a blue (red) dashed line for the (zero)-field-cooled
system. Coordinates are equivalent modulo $a_{0}$.}
\label{fig:high-fields}
\end{figure}

In the discussion above, we have analyzed the interaction of pinning centers
with a single flux line. In high fields, where the vortex separation $a_{0}$
is comparable to the maximal pinning length $x_{+} \sim \kappa \xi$, this
picture needs to be modified as the periodicity of the pinning potential has
to be properly accounted for. In the vicinity of $H_{c2}$, the pinning
potenial is dominanted by the lowest harmonic, $e_{p}(x) \propto [1-\cos(2\pi
x / a_{0})]$, and the corresponding force takes the form $f_{p}(x) \approx
-f_{0}\sin(2 \pi x / a_{0})$. Analyzing the characteristic lengths for the
present situation, one finds that (i) the position $x_{m}$ of steepest slope
in $f_{p}$ coincides with $a_{0}/2$, (ii) $x_{0\mathrm{ \scriptscriptstyle L}}
= x_{m} - f_{p}(x_{m})/\bar{C} = x_{m}$ because $f_{p}(x_{m}) = 0$, and (iii)
$x_{0}(T) = x_{0\mathrm{\scriptscriptstyle L}}$ because of symmetry arguments.
Furthermore, the branch edges $\pm x_{-}$ have disappeared and those at $\pm
x_{+}$ overlap with the next period, i.e., $|x_{+}| > a_{0}/2$, see Fig.\
\ref{fig:high-fields}. As a result, in a zero-field-cooled sample, the
high-field limit of the Campbell length $\lambda_{\mathrm{\scriptscriptstyle
C}}$ is always determined by the (single) jump at $a_{0}/2$, $\Delta \fpin =
\Delta \fpin|_{a_{0}/2}$; in analogy to our previous nomenclature, we call
this the case (a') and note that this reversible (non-hysteretic) behavior
becomes the generic case at high fields. In the zero-field-cooled (or Bean
critical) state, the penetration depth $\lambda_{\mathrm{\scriptscriptstyle
C}}$ involves the (slightly smaller) jump at $x_{+}$, $\Delta \fpin = \Delta
\fpin|_{x_{+}}$, and hence $\lambda_{\mathrm{\scriptscriptstyle
C}}|_{\mathrm{\scriptscriptstyle FC}} \lesssim \lambda_{\mathrm{
\scriptscriptstyle C}}|_{\mathrm{\scriptscriptstyle ZFC}}$.

Again, simple closed-form expressions can be found at small values of $\kappa
\gtrsim 1$. Near the Labusch point, the results Eqs.\ \eqref{eq:fpin_x0L} and
\eqref{eq:fpin_xpm} remain valid [with $\gamma = (4\pi^3/a_0^3) f_0$ and
$x_{0\mathrm{\scriptscriptstyle L}} \to a_0/2$], from which expressions for
$\lambda_{\mathrm{\scriptscriptstyle C}}$ follow immediately. At large
$\kappa$, the low-field result $\Delta f_\mathrm{pin} \simeq \C \kappa \xi$ is
cut off by the lattice period when $\kappa \xi > a_0$ and hence $\Delta
f_\mathrm{pin} \simeq \C a_0$, resulting in a Campbell length 
\begin{align}\label{eq:lambda_ssp}
   \lambda_{\mathrm{\scriptscriptstyle C}}^{2} &\simeq
      \frac{B_{0} \Phi_{0}}{4\pi n_{p} t_{\perp}} \frac{1}{\C a_{0}}
      \sim \frac{\lambda^{2}}{n_{p} a_{0} \xi^{2}}\frac{\xi}{a_{0}},
\end{align}
a factor $\kappa\xi/a_0$ larger than the low-field result
\eqref{eq:lambda_est}. However, for such strong pinning the applicability of
the elastic theory becomes questionable. Indeed, as suggested by numerical
analysis \cite{Schoenenberger1996, KoshelevPrivComm}, the elastic theory might
break down due to plastic instabilities of the vortex lattice. It has been
shown \cite{Schoenenberger1996} that a vortex detaches from an infinitely
strong pinning center via loop formation and subsequent vortex cutting and
reconnection, which is a highly non-elastic process. Similarly, a
computational study based on a time-dependent Ginzburg-Landau solver
\cite{Sadovskyy2015a,KoshelevPrivComm} has demonstrated that small pinning
centers are not capable of holding multiple vortices at the same time.
Indeed, rather than trapping a second flux line, the defect trades one vortex
for the next, with the first vortex pushed out of the pinning well by the
following one. These insights suggest that the maximal pinning distance
$x_{+}$ should be limited to $a_0$, producing a force jump $\Delta
f_\mathrm{pin} \simeq \C a_0$. This result coincides with the one above and
the Campbell length is still given by Eq.\ \eqref{eq:lambda_ssp}. The
situation is more subtle when considering the value of $j_c$ at high fields
and strong pinning. Assuming the elastic theory to remain valid, one obtains a
jump in energy $\Delta \epin = f_{0} a_{0}$ and $j_c$ is reduced by a factor
$a_0/ \kappa \xi < 1$ as compared to the low-field result \eqref{eq:jc_est}
(note that multiply overlapping pinned branches appear when $\kappa \xi >
a_0$). Accounting for plasticity, the jump in energy is even further reduced,
$\Delta \epin \approx \C a_{0}^{2} / 2$, and the critical current takes the
universal form $j_{c} \sim j_{\mathrm{dp}} (n_{p} a_{0} \xi^{2})$, a factor
$(a_{0}/\kappa\xi)^2$ smaller than the low-field result \eqref{eq:jc_est}.

For mid-range magnetic fields, neither the single-vortex nor the sinusoidal
force profile is accurate. Starting from the limit of high fields, in addition
to the basic sinusoidal force profile, further higher-order harmonics need to
be taken into account. As the field is sufficiently lowered, the position
$x_{m}$ of the maximal slope in the (bare) force profile detaches from
$a_{0}/2$, and a second (unpinned) branch develops. 

\subsubsection{Comparison between different regimes}\label{sec:comp}

\begin{figure}[tb] \includegraphics[width=6.0cm]{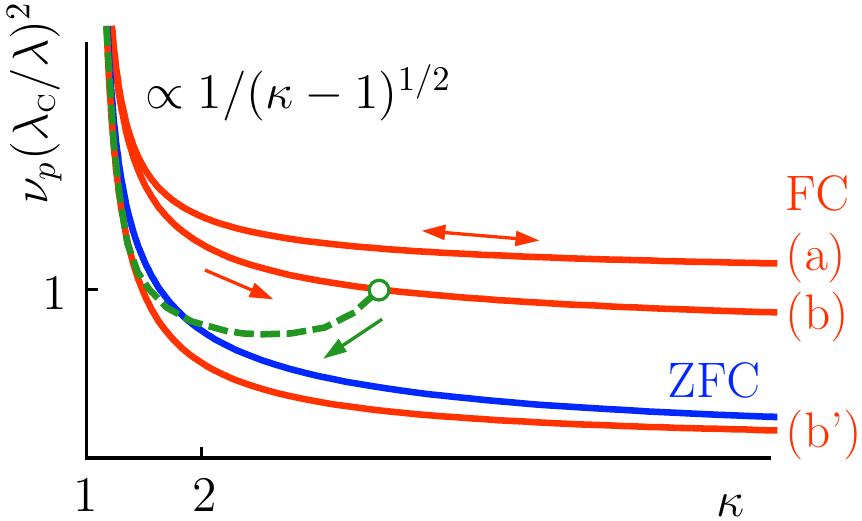}
\caption{Schematic view of the $\kappa$-dependence of the parameter $\nu_{p}
(\lambda_{\mathrm{\scriptscriptstyle C}} / \lambda)^{2}$, with $\nu_{p} =
n_{p} a_{0} \xi^{2} \ll 1$ the small dimensionless density parameter. Upon
crossing the Labusch point $\kappa = 1$ and entering the strong pinning
regime, all curves first decrease as $1/(\kappa - 1)^{1/2}$. Subsequently, the
cases FC (a) and FC (b) decay weakly $\propto \xi/t_\perp$ and $\propto
\xi^2/t_\perp x_-$, respectively. The cases ZFC and FC (b') decay more rapidly
$\propto \xi/\kappa t_\perp$. At large $\kappa$, the decrease slows down when
$x_+ \simeq \kappa \xi$ and $x_-$ reach the scale $a_0$; note that $t_\perp =
2 x_-$ within our analysis.  The field-cooled cases (b) and (b') exhibit a
hysteretic response upon reversing the direction of $\kappa$ (see dashed
line). The hysteresis loop shown for case FC (b) merges with the curve FC (b')
when taking $\kappa$ back to unity. Similarly, reversing $\kappa$ on curve FC
(b') one approaches the curve FC (b) (not shown).}
\label{fig:comp}
\end{figure}

It is interesting to analyze the scaling behavior of the Campbell length as a
function of the strong pinning parameter $\kappa$. Indeed, when expressing
$\lambda_{\mathrm{\scriptscriptstyle C}}^{2}$ [\eq \eqref{eq:Campbell}] in
units of $\lambda^{2} / \nu_p$, with $\nu_p = n_{p} a_{0} \xi^{2}$ the
dimensionless small density parameter, we find
\begin{align}\label{eq:characteristic-scaling}
   \nu_p \frac{\lambda_{\mathrm{\scriptscriptstyle C}}^{2}}
        {\lambda^{2}} 
      &\sim \frac{\C \xi}{\Delta \fpin}\frac{\xi}{t_\perp}
\end{align}
where we have used that $\C \sim \varepsilon_0/a_0$, see Sec.\
\ref{sec:formalism}. When pushing the system across the Labusch point $\kappa
= 1$ into the strong pinning regime, we find a universal scaling [see Eqs.\
\eqref{eq:fpin_x0L} and \eqref{eq:fpin_xpm}]
\begin{align}\label{eq:characteristic-scaling-near-Labusch}
   \nu_p \frac{\lambda_{\mathrm{\scriptscriptstyle C}}^{2}}
        {\lambda^{2}} \sim \frac{1}{\sqrt{\kappa - 1}},
\end{align}
which is valid at all fields and for different vortex states (FC as well as
ZFC), see Fig.\ \ref{fig:comp}. Combining this result with the standard
scaling \cite{Blatter2004} of the critical current density $j_{c} \sim
j_{\mathrm{dp}} \nu_{p} (\xi / a_{0})^{2} (\kappa - 1)^{2}$ in the vicinity of
the Labusch point, we arrive at the relation $j_{c} \sim (c \xi B_{0} /
\lambda_{\mathrm{\scriptscriptstyle C}}) (\kappa - 1)^{3/2}$ which strongly
differs from the scaling $j_{c} \sim c \xi B_{0} / \lambda_{\mathrm{
\scriptscriptstyle C}}$ obtained within a phenomenological approach.  At
intermediate values of $\kappa$, we can write $\Delta f_\mathrm{pin} \sim \C
x_\mathrm{jp}$ with $x_\mathrm{jp} = x_-,\ x_{0\mathrm{\scriptscriptstyle L}}
\simeq \xi,\ x_+$ for the cases FC (b), FC (a), and ZFC/FC (b'). For the cases
FC (b) and (a) the further change in $\nu_p \lambda_{\mathrm{
\scriptscriptstyle C}}^{2}/ \lambda^{2}$ is small, $\propto \xi / t_{\perp}$
and $\propto \xi^2 / t_\perp x_{-}$, respectively, while a pronounced decrease
appears for the ZFC and FC (b') cases, $\propto \xi/\kappa t_\perp$.  At large
$\kappa \gg 1$, the quantities $x_-,\ x_+,\ t_\perp$ saturate as they reach
the scale $a_0$, with a corresponding change in the expression for $\nu_p
\lambda_{\mathrm{\scriptscriptstyle C}}^{2}/ \lambda^{2}$.  Finally, case FC
(b') assumes that $x_+ \simeq \kappa \xi$ decreases with increasing $\kappa$
and naturally terminates when this condition is violated.  The scaling
behavior of $\nu_p \lambda_{\mathrm{\scriptscriptstyle C}}^{2}/\lambda^{2}$
and the appearance of hysteretic behavior upon reheating is illustrated in
Fig.\ \ref{fig:comp}. The low-field scaling discussed above changes over to
the high-field behavior (see section \ref{sec:high-fields}, $\nu_p
\lambda_{\mathrm{\scriptscriptstyle C}}^{2}/\lambda^{2} \propto
1/\sqrt{\kappa-1},\ 1$ at small and large $\kappa > 1$, respectively) when the
intervortex distance $a_{0}$ approaches $\xi$.  These results can be used to
characterize the pinscape by combining theoretical input on $\kappa(T,H)$ for
various pinning models (see below) with experimental data for
$\lambda_{\mathrm{\scriptscriptstyle C}}$. Such information is of great value
when simulating vortex dynamics within a numerical approach, e.g., using
time-dependent Ginzburg-Landau theory\cite{Sadovskyy2015a,Sadovskyy2015b}.
\section{Pinning Models}\label{sec:specific-pinning-models}
In order to proceed further towards quantitative predictions of the Campbell
response, we have to specify all relevant quantities entering the pinning
problem. In particular, the temperature- and field-dependence of the vortex
elasticity $\C$ and the pinscape energy $e_{p}$ have to be determined.  This
is the central topic of this Section and will allow us later to follow the
evolution of the effective force $\fpin$ upon cooling and reheating the system
in the field as typically done in an experiment. Here, we focus on the
comparison between different pinning models and defer the comparison to
experiments to \sect \ref{sec:experiments}.

The $T$- and $H$-dependence of the effective elasticity $\bar{C}$ and of the
pinning energy $e_{p}$ is mainly determined by the Ginzburg-Landau parameters
$\lambda$ and $\xi$, with the superfluid density $n_s \propto \lambda^{-2}$
scaling both with temperature and field, $\lambda^{-2} \simeq
\lambda^{-2}_{0}(1-T/T_c)[1-B/H_{c2}(T)] \simeq \lambda^{-2}_{0}
(1-\tau-b_{0})$, while the coherence length scales with temperature only,
$\xi^{-2}(T) = \xi_0^{-2}(1-\tau)$. Here we have used the scaling of the upper
critical field $H_{c2}(T) = \Phi_0/2\pi\xi^2(T)$ and have introduced the
reduced temperature $\tau = T/T_{c}$ and the reduced field $b_{0} =
B_{0}/H_{c2}(0)$. Note that the superfluid density vanishes on approaching
$H_{c2}(T)$ where $1-\tau-b_{0} = 0$. In this section, we shall not burden our
expressions with the more complicated details of the $H$- and $T$-dependence
in the superconducting phase but rather extrapolate the Ginzburg-Landau
scaling valid near $T_{c}$ to the entire phase diagram. In \sect
\ref{sec:experiments}, where we confront our predictions with experimental
data, a more accurate scaling will be chosen.

The Labusch parameter is given through the ratio of pinning curvature and
elasticity, $\kappa = \max_{x}[-e_{p}''(x)]/\C$. For small defects, we
approximate the curvature $-e_{p}'' \approx e_0 V_{\mathrm{pin}}/\xi^2$, with
$e_0$ the typical gain in energy density and $V_{\mathrm{pin}}$ the relevant
pinning volume, hence
\begin{align}\label{eq:kappa-approx}
   \kappa \approx \frac{e_{0} V_{\mathrm{pin}}}{\xi^{2}\C}.
\end{align}
In the following, we consider four different pinning models (metallic and
insulating inclusions, as well as $\delta T_{c}$-, or $\delta \ell$-pinning)
and evaluate the behavior of $\kappa$ within the $H$-$T$ phase diagram.

\subsection{Elasticity $\vec{\C}$}\label{sec:elasticityC}
We first evaluate the effective elasticity $\C$, a quantity that is
independent of the chosen pinning model. Starting from its definition
\eqref{eq:Cbar-def} and the subsequent discussion, the effective elasticity
$\C = (\nu a_{0}^{2}/\lambda)[c_{66} c_{44}(\vec{0},0)]^{1/2}$ involves the
non-dispersive shear ($c_{66}$) and the bulk tilt [$c_{44}(0)$] moduli as well
as the London penetration depth $\lambda$. Inserting the standard
expressions\cite{Brandt1986,Blatter1994} for the elastic moduli
\begin{align}\label{}
   c_{44}(\vec{0},0) &= \frac{B_{0}^{2}}{4\pi}
   \!\!&\mathrm{and}\;\;
   c_{66} &\simeq \frac{\varepsilon_{0}}{4 a_{0}^{2}}
    \Big[1-\frac{B_{0}}{H_{c2}(T)}\Big]
\end{align}
and using the scaling $\varepsilon_0 \propto (1-\tau - b)$ while
$c_{66} \propto (1-B_0/H_{c2})^2(1-T/T_c) = (1-\tau-b)^2/(1-\tau)$ we 
find that
\begin{align}\label{eq:C-exp}
   \C &= \nu \bigg(\frac{\Phi_{0}}{4 \pi \lambda_{0}}\bigg)^{2}
   \sqrt{\frac{b_{0}}{2\xi_{0}^{2}}} \frac{(1-\tau-b_{0})^{3/2}}
   {(1-\tau)^{1/2}}.
\end{align}
The factor $(1-\tau-b_{0})^{3/2}$ describes the softening of the lattice near
the $H_{c2}$-line.

\subsection{Small defects}\label{sec:general-approach}
We consider a defect in the form of a small inclusion of radius $\rho \ll
\xi$. A vortex placed a distance $x$ away from this pin will experience an
energy decrease
\begin{align}\label{eq:pinning-energy-def}
   e_{p}(x) = - \int dx' e_{0}(x')\, [1 - |\psi_{0}(x-x')|^2].
\end{align}
The shape of the vortex solution $\psi_0(x)$ can be obtained within
Ginzburg-Landau (GL) theory\cite{Ginzburg1950}; at low fields, it is well
described by the expression\cite{Schmid1966, Clem1975} $|\psi_{0}(x)|^2 =
x^{2}/(x^{2} + 2\xi^{2})$, producing a Lorentzian shape for the pinning
potential
\begin{align}\label{eq:ep-low}
   e_{p}(x) = -\frac{e_{0} V_{\mathrm{pin}}}{1 + x^{2}/2\xi^{2}}.
\end{align}
In the high field limit $1 - B_{0}/H_{c2} \ll 1$, we approximate the vortex
solution \cite{SaintJames1969} by the one-dimensional harmonic
\begin{align}\label{eq:psi-high}
   |\psi_{0}(x)|^2 = \frac{1}{2}[1-\cos(2\pi x/a_{0})],
\end{align}
and evaluating \eq \eqref{eq:pinning-energy-def}, we arrive at a periodic
pinning profile
\begin{align}\label{eq:ep-high}
   e_{p}(x) = - \frac{e_{0} V_{\mathrm{pin}}}{2}[1 + \cos(2\pi x/a_{0})].
\end{align}
\begin{figure}[tb]
\includegraphics[width=7.0cm]{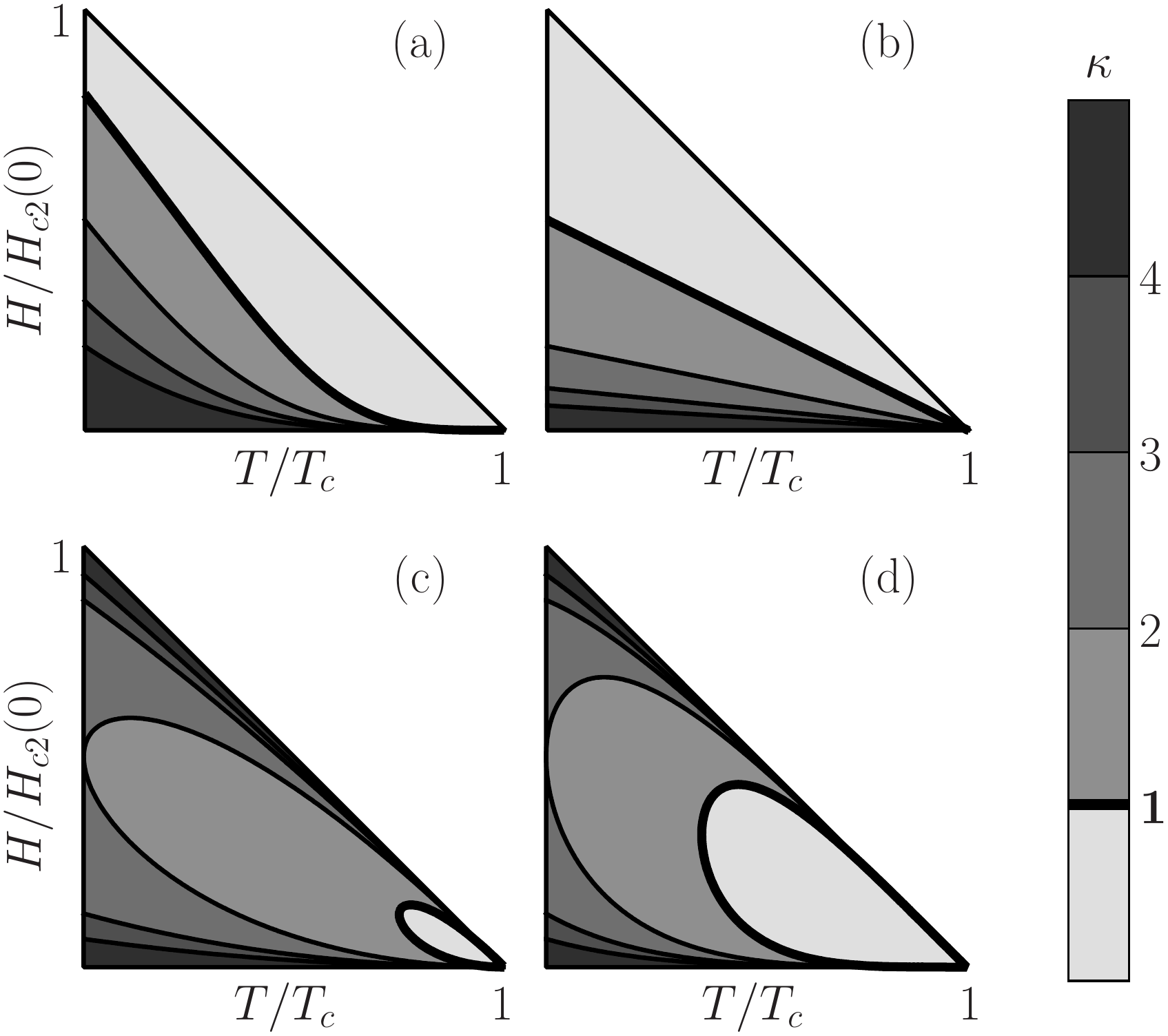}
\caption{Generic density plot of the strong pinning parameter $\kappa(T,H)$
for four different pinning models. The Labusch criterion $\kappa = 1$
determines the transition line (bold), where the pinscape changes from weak ($\kappa
< 1$) to strong ($\kappa > 1$) pinning. Depending on the parameters, this
separatrix assumes a different shape and location in the diagram. For
insulating (a) and metallic (b) inclusions the pinscape is weak upon crossing
$H_{c2}(T)$ and turns strong at low fields/temperatures, see \eqs
\eqref{eq:kappains} and \eqref{eq:kappamet}, respectively. For defects
inducing a local change in $T_{c}$ (c) or in the mean free path $\ell$ [via a
local change of the effective mass, panel (d)] the pinscape is strong upon
entering the superconducting phase, see \eqs \eqref{eq:kappadTc} and
\eqref{eq:kappadell}, respectively. For a good visibility of all the features,
we have used the parameters $\rho^{3} = 2 \xi_{0}^{3}$ and $\delta T_{c} /
T_{c} = \delta \ell/\ell = 0.5$.}
\label{fig:kappa}
\end{figure}

\subsection{Insulating defect}\label{sec:insulating-pin}
For an insulating inclusion, the typical energy that a vortex state gains by
aligning a flux line with the defect is determined by the condensation energy
density and the pin volume. The former derives from
a minimization of the Ginzburg-Landau\cite{Ginzburg1950} (GL) functional
$f_{\mathrm{\scriptscriptstyle GL}} = \alpha |\psi|^2 + \beta|\psi|^4/2$,
providing an order parameter $|\Psi_0|^2 = |\alpha|/\beta$; inserting this
back to $f_{\mathrm{\scriptscriptstyle GL}}$, we obtain $e_0 =
-|\alpha|^2/2\beta$. While in standard GL theory the temperature dependence
derives from $\alpha(T) = \alpha_0 (T-T_c) = -\alpha_0 T_c (1-\tau)$, close to
the $H_{c2}(T)$-line we can adopt a lowest Landau level approximation of the
GL functional \cite{Ruggeri1976} by replacing this temperature dependence with
a temperature and field dependence $\alpha(T,B) = -\alpha_0 T_c
(1-\tau-b_{0})$ (the order parameter then assumes the role of the amplitude of
the space-modulated solution). The combination with the expression for the
(zero-temperature) thermodynamic critical field $H_{c0}^2/4\pi = \alpha_0^2
T_c^2/\beta$ then produces an overall temperature and field scaling of the
condensation energy of the form
\begin{align}\label{eq:eins}
   e_{0}(T,B_{0})
         \approx \frac{H_{c0}^{2}}{8\pi}(1-\tau-b_{0})^{2},
\end{align}
in agreement with the discussion in \sbonlinecite{DeGennes1966}. Combining
this result with the effective elasticiy $\bar{C}$ in \eq \eqref{eq:C-exp}, we
make use of \eq \eqref{eq:kappa-approx} to find the following explicit
dependence on $b_0$ and $\tau$ of the strong pinning parameter,
\begin{align}\label{eq:kappains}
   \kappa
   \approx
   \frac{\rho^{3}}{\xi_{0}^{3}} \sqrt{\frac{1}{b_{0}}}(1-\tau)^{3/2}
   (1-\tau-b_{0})^{1/2}.
\end{align}
For insulating pins the strong pinning parameter reveals two important
asymptotic regimes, see \fig \ref{fig:kappa}(a). First, $\kappa$ vanishes
along the entire $H_{c2}$-line defined through $1 - \tau - b_{0} = 0$. As a
consequence, the insulating defects act as weak pins upon crossing
$H_{c2}(T)$. Second, $\kappa$ grows as $b_{0}^{-1/2}$ at low fields (a
consequence of the softening of $\C$ at low fields), thus guaranteeing that the
defect turns into a strong pinning center with $\kappa > 1$.

\subsection{Metallic defect}\label{sec:metallic-pin}
Similar happens in the case of a metallic defect which affects the
superconductor via the proximity effect. For a metallic inclusion, the order
parameter is substantially suppressed within a volume $\sim \xi^{3}$ around
the pin. This provides the possibility for the flux line lattice to gain the
condensation energy density \eqref{eq:eins} over a larger volume
$V_{\mathrm{pin}} \to V_{\mathrm{pin}}^{\mathrm{eff}} \approx (4\pi/3)\xi^{3}$.
In this case, the real size of the defect drops out of the final result and the
strong pinning parameter
\begin{align}\label{eq:kappamet}
   \kappa \approx \sqrt{\frac{1}{b_{0}}}(1-\tau-b_{0})^{1/2}
\end{align}
shows a qualitatively similar behavior as the insulating pin in \eq
\eqref{eq:kappains}, see \fig \ref{fig:kappa}(b): the metallic pins are weak
upon entering the superconducting phase when traversing the $H_{c2}$-line and
they turn strong at low temperatures and fields where $\kappa \approx
b_{0}^{-1/2}$.

\subsection{$\vec{\delta T_{c}}$-pinning}\label{sec:deltaTc-pin}
A very different pinning behavior is observed when defects locally change the
critical temperature, $T_{c} \to T_{c} - \delta T_{c}$. Such a local variation
in $T_c$ has to be included in the quadratic term of the GL energy functional
and  adds a correction $\alpha_0 \delta T_c |\psi|^2$. Making use of the
above results for $|\psi|^2$ and $H_{c0}^2$, we arrive at the following
expression for the local energy gain
\begin{align}\label{eq:edeltaTc}
   e_{0}(T,B_{0})
   \approx \frac{H_{c0}^{2}}{4\pi} \frac{\delta T_{c}}{T_{c}} 
                 (1 - \tau - b_{0}).
\end{align}
The combination with the expression \eqref{eq:C-exp} for the elasticity $\C$
provides the scaling for $\kappa$ in the form
\begin{align}\label{eq:kappadTc}
   \kappa \approx \frac{\rho^{3}}{\xi_{0}^{3}}
                  \frac{\delta T_{c}}{T_{c}}
                  \sqrt{\frac{1}{b_{0}}}(1-\tau)^{3/2}(1-\tau-b_{0})^{-1/2}.
\end{align}
As a result, $\kappa$ is divergent both at low fields $b_{0} \to 0$ and near
the $H_{c2}(T)$-line. On the other hand, the factor $(1-\tau)^{3/2}$
suppresses $\kappa$ near $T_{c}$. This leads to a peculiar
weak-pinning lobe extending from $(T_{c}, H = 0)$ into the phase diagram, see
\fig \ref{fig:kappa}(c). Lowering the temperature at a constant field
(horizontal cut) or changing the field at a constant temperature (vertical
cut), the system might cross the Labusch point twice, changing
from strong to weak and back to strong pinning. As observed by Larkin and
Ovchinnikov \cite{Larkin1979}, the strengthening of pinning
near $H_{c2}(T)$ manifests itself in a sharp increase of the critical
current, a feature known as peak effect \cite{Autler1962, DeSorbo1964}.

\subsection{$\vec{\delta\ell}$-pinning}\label{sec:deltaell-pin}
Finally, we address the pinning due to local changes of the mean free path,
$\ell \to \ell - \delta \ell$. The dependence of the Ginzburg-Landau
functional on the mean free path appears in the gradient term $(\hbar^2/2m)
|\nabla\psi|^2$. Indeed, a microscopic calculation\cite{Gorkov1959} provides
the additional factor $\chi(\rho_{\ell})$, $\rho_{\ell} = \hbar
v_\mathrm{\scriptscriptstyle F}/2 \pi T_c \ell \simeq \xi_0/\ell$, with $\chi
\approx 1$ and $\chi \approx [\pi^2/7\zeta(3)]/\rho_{\ell}$ in the clean and
dirty limits, respectively. As a result, the coherence length $\xi(T)$
depends on disorder via
\begin{align}\label{eq:xi_ell}
   \xi^2 \approx \xi_0^2 \chi(\rho_{\ell})/(1-\tau)
\end{align}
with $\xi_0$ the $T=0$ clean-limit coherence length. The (quenched)
fluctuations in $\ell$ translate into fluctuations in the gradient term and
entail a change in the energy density $e_{0}$ of the form
\begin{align}\label{eq:edeltaM}
   e_{0}(T,B_{0})
    \approx \frac{H_{c0}^{2}}{4\pi} \frac{\delta \ell}{\ell} 
    (1-\tau) (1 - \tau - b_{0}).
\end{align}
We then arrive at a Labusch parameter in the form
\begin{align}\label{eq:kappadell}
   \kappa \approx \frac{\rho^{3}}{\xi_{0}^{3}} \frac{\delta \ell}{\ell}
   \sqrt{\frac{1}{b_{0}}}(1-\tau)^{5/2}(1-\tau-b_{0})^{-1/2},
\end{align}
exhibiting a qualitative similar behavior as the one found for $\delta
T_{c}$-pinning but with a larger exponent 5/2 for the $(1-\tau)$ factor
(3/2 for $\delta T_{c}$-pinning), pushing the weak-pinning lobe deeper into
the phase diagram, see \fig \ref{fig:kappa}(d).

\section{Comparison to Experiments}\label{sec:experiments}
Equipped with a microscopic expression for the Campbell penetration depth, we
discuss experimental signatures that provide strong support for our new
results. Below, we focus on few original studies by Campbell
\cite{Campbell1969,Campbell1971} and Lowell \cite{Lowell1972_1,Lowell1972_2}
as well as more recent studies by Prozorov and co-workers, see Refs.\
\sbonlinecite{Prozorov2003}, \sbonlinecite{Kim2013}, and
\sbonlinecite{Gordon2013}.

\subsection{General comparison}\label{sec:general-comparison}

We have identified four major experimental signatures that find a natural
explanation within our analysis of $ac$ magnetic response.

\paragraph{Low versus high $dc$ fields.} 
In early work, e.g., by Campbell \cite{Campbell1969} or Lowell
\cite{Lowell1972_1}, it has been noted that the $ac$ magnetic response does
not depend on the state preparation (field-cooled or zero-field-cooled). A
simple (piece-wise linear) force model was put forward
\cite{Campbell1971,Lowell1972_2} in support of this result. The dependence of
the $ac$ magnetic penetration depth $\lambda_{\mathrm{ \scriptscriptstyle C}}$
on the vortex state preparation was first reported by Prozorov and
co-workers\cite{Prozorov2003}. In recent years, different Campbell lengths for
the field-cooled and zero-field-cooled states have been observed
\cite{Kim2013,Gordon2013} in a wide range of materials, including Niobium,
MgCNi$_{3}$, SrPd$_{2}$Ge$_{2}$, the high-temperature superconductor
Bi$_{2}$Sr$_{2}$CaCu$_{2}$O$_{8}$, Pr$_{1-x}$Ce$_{x}$CuO$_{4}$, and the
organic superconductor $\beta''$-(ET)$_{2}$SF$_{5}$CH$_{2}$CF$_{2}$SO$_{3}$.
The new results provided by our strong pinning analysis are compatible with
both types of observations: at high $dc$ fields, the typical setup of early
experiments, the Campbell lengths are (almost) identical, see \sect
\ref{sec:high-fields}, while they are different (sometimes even
parametrically) at low fields, see sections \ref{sec:zfc} and \ref{sec:fc}.

\paragraph{Finite $\lambda_{\mathrm{\scriptscriptstyle C}}$
in the critical state.}
The phenomenological theory, on which the interpretation of most $ac$
experiments has been based so far, predicts\cite{Prozorov2003} a divergent
Campbell length for the zero-field-cooled state,
$\lambda_{\mathrm{\scriptscriptstyle C}} \propto (1-j/j_{c})^{-1/4}$, as the
curvature $\alpha(j) \propto (j_{c}-j)^{1/2}$ of the pinning well vanishes on
approaching the critical state. Not only is the experimentally observed
Campbell length in the Bean critical state finite, but in some materials it is
even smaller than that of the field-cooled state,
$\lambda_{\mathrm{\scriptscriptstyle C}}|_{\mathrm{\scriptscriptstyle ZFC}}
<\lambda_{\mathrm{\scriptscriptstyle C}}|_{\scriptscriptstyle \mathrm{FC}}$.
Both features are well understood within the strong pinning framework. The
Campbell length $\lambda_{\mathrm{\scriptscriptstyle C}}$ results from an
averaging of the local curvature which can (depending on the pinning
parameters) get reduced when changing the branch occupation from the
field-cooled state to the zero-field-cooled state. In the latter situation,
the application of an $ac$ field will first generate flux pulses that
penetrate the sample and change the $dc$ field inside the
material\cite{Willa2015b}. At the end of this transient initialization, the
response of the vortex system is perfectly regular and characterized by a
finite Campbell length $\lambda_{\mathrm{\scriptscriptstyle C}}$.

\paragraph{Hysteresis upon thermal cycling of
$\lambda_{\mathrm{\scriptscriptstyle C}}|_{\mathrm{\scriptscriptstyle FC}}$.}
The strong pinning framework of $ac$ magnetic response predicts the appearance
of hysteretic Campbell lengths for the field-cooled samples upon thermal
cycling. Such hysteretic behavior has been observed in experiments by
Prozorov and co-workers, see Ref.\ \sbonlinecite{Willa2015a}.

\paragraph{Universality of $\lambda_{\mathrm{\scriptscriptstyle C}}|_{\mathrm{
\scriptscriptstyle ZFC}}$ for different critical states.}
Within our microscopic analysis, the direction of the Lorentz force $\pm
j_{c}$ does not affect the asymptotic (i.e., large times $t \gg 2\pi /
\omega$) oscillatory response of the vortex lattice in the critical state.
Hence, the Campbell length is independent on whether the external field $H$ is
reached from below (ramping up) or from above (ramping down). This
independence has been experimentally demonstrated \cite{Prozorov2003}. We
note that the transient behavior before reaching the asymptotic periodic
regime may exhibit differences between the two ramping directions, as an
opposite $dc$-shift is expected when ramping the field down to $H$, with the
number of cycles needed to reach the asymptotic behavior depending on the
depth of the critical state. This prediction could be verified in an
experiment.

\subsection{Comparison to SrPd$_{2}$Ge$_{2}$}

\begin{figure}[tb]
\includegraphics[width=.48\textwidth]{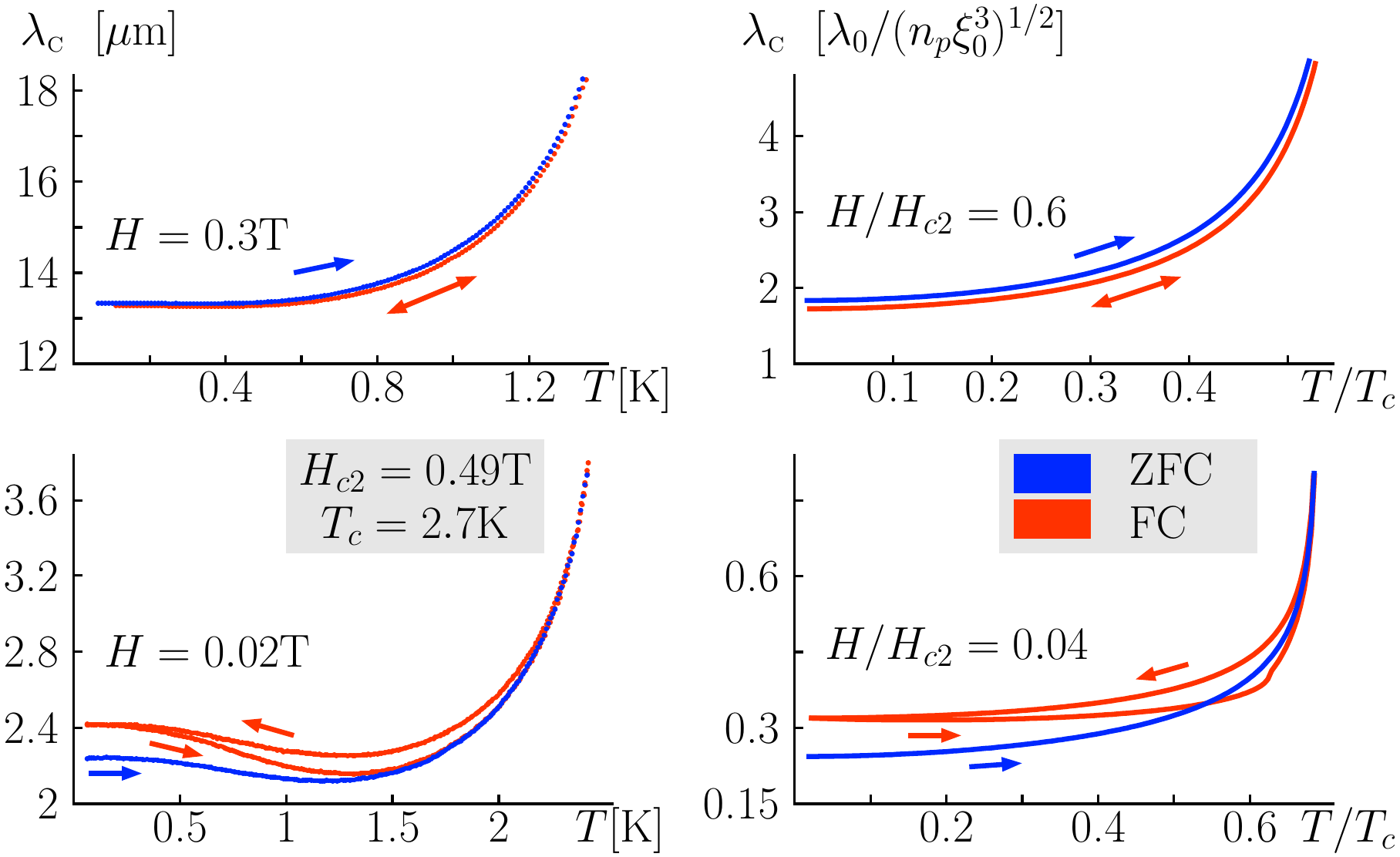}
\caption{
Comparison between the Campbell length obtained from experiments (left) and
from numerics (right). On the experimental side, we show representative traces
for high ($H = 0.3\mathrm{T} \approx 0.6 H_{c2}$) and low fields ($H =
0.02\mathrm{T} \approx 0.04 H_{c2}$), see Ref.\ \sbonlinecite{Kim2013} and
\sbonlinecite{Willa2015a}. The corresponding theoretical curves are obtained
from (i) solving \eq \eqref{eq:Labusch-equation-1d} numerically for an
insulating pin (see \sect \ref{sec:insulating-pin}), (ii) determining the
relevant jumps in the force profile, and (iii) evaluating
$\lambda_{\mathrm{\scriptscriptstyle C}}$ through \eq \eqref{eq:Campbell}.
The sharp upturn appearing at $T \approx 0.65 T_{c}$ upon reheating the
field-cooled state in low fields is due to the change in the jump position
$x_{\mathrm{jp}}$ as it reaches the branch edge $x_{+}$ (corresponding to
$T_{2}$ in Fig.\ \ref{fig:hysteresis}).}
\label{fig:comp-to-exp}
\end{figure}

Finally, we provide a semi-quantitative comparison of our microscopic analysis
of the Campbell length $\lambda_{\mathrm{\scriptscriptstyle C}}$ with
measurements on a single-crystal germanide superconductor SrPd$_{2}$Ge$_{2}$
with $T_{c} = 2.7~\mathrm{K}$ and $H_{c2} = 0.49~\mathrm{T}$. Vortex pinning
in this ternary compound, parent to the iron- and nickel-pnictides, is likely
to be strong.\cite{Sung2012,Kim2013} Its $ac$ response has been investigated
with a tunnel-diode technique in Ref.\ \sbonlinecite{Kim2013}. We focus on two
traces of $\lambda_{\mathrm{\scriptscriptstyle C}}$ recorded at different
applied $dc$ fields 0.02 T and 0.3 T and taken from Fig.\ 4(a) of Ref.\
\sbonlinecite{Kim2013}. In Fig.\ \ref{fig:comp-to-exp} (left) we show an
enlarged view of these traces, with the zero-field-cooled data in blue and the
thermally cycled field-cooled data (see arrows for the temperature direction)
in red. At high fields $0.3~\mathrm{T} \approx 0.6 H_{c2}$, the Campbell
lengths are almost identical, but with $\lambda_{\mathrm{\scriptscriptstyle
C}}|_{\mathrm{\scriptscriptstyle ZFC}}$ slightly larger than
$\lambda_{\mathrm{\scriptscriptstyle C}}|_{\mathrm{\scriptscriptstyle FC}}$.
The low-field trace at $0.02~\mathrm{T} \approx 0.04 H_{c2}$ is much richer:
the field-cooled and zero-field-cooled Campbell lengths clearly differ from
each other. Moreover, the Campbell length of the field-cooled state shows a
strong hysteresis upon thermal cycling. The heating branch (arrow to the
right) deviates from the cooling branch and approaches the zero-field-cooled
Campbell length at higher temperatures. Finally all Campbell length curves
feature a minimum at around $1.2~\mathrm{K}$.

The theoretical analysis of $\lambda_{\mathrm{\scriptscriptstyle C}}$, see
right of Fig.\ \ref{fig:comp-to-exp}, requires the knowledge of the pinscape
$e_{p}(x)$ for all fields $B_{0}$. We interpolate between the two limits of
low [\eq \eqref{eq:ep-low}] and high [\eq \eqref{eq:ep-high}] fields by
periodically summing the low-field profile,
\begin{align}\label{}
   \chi(x) = \sum_{n = -\infty}^{\infty} \frac{1}
      {1 + (x + n a_{0})^{2}/2\xi^{2}}
\end{align}
and applying the normalization (neglected in Ref.\ \sbonlinecite{Willa2015a})
\begin{align}\label{}
   e_{p}(x) = e_{0} V_{\mathrm{pin}} \frac{\chi(x) 
    - \chi(0)}{\chi(a_{0}/2) - \chi(0)}.
\end{align}
For a better comparison to experiment, we replace the simple Ginzburg-Landau
scaling $1-\tau$ used in \sect \ref{sec:specific-pinning-models} by the more
accurate two-fluid-model scaling $1-\tau^{2}$; the latter properly captures
the saturation of the phenomenological parameters at low temperatures.  In
order to find the temperature and field dependence of $\lambda_{\mathrm{
\scriptscriptstyle C}}$, we numerically evaluate the force profile
$f_\mathrm{pin}(x)$ and the relevant jumps $\Delta f_\mathrm{pin}$ for the
zero-field-cooled and field-cooled situations. For the latter, we make use of
a numerical routine that follows the branch evolution and the associated
occupation upon lowering the system's temperature and its subsequent heating.

The quantity $(n_p\xi_0^3)^{1/2} \lambda_{\mathrm{\scriptscriptstyle C}} /
\lambda_0$ solely depends on the ratio $\rho/\xi_0$, with $\rho$ the radius of
the insulating defect. This parameter governs the extent of the strong pinning
region within the $H$-$T$ diagram.  It turns out, that choosing an insulating
defect of radius $\rho \approx 1.82\, \xi_{0}$ ($\rho \approx 1.6 \, \xi_{0}$
in Ref.\ \sbonlinecite{Willa2015a}), see \eq \eqref{eq:eins}, provides a good
description of the experimental data.  Indeed, the results shown on the right
of \fig \ref{fig:comp-to-exp} reproduce all relevant features of the
experiment: These are the closeness between the ZFC and FC Campbell lengths at
high fields, as well as their sequence in magnitude, $\lambda_{\mathrm{
\scriptscriptstyle C}}|_{\mathrm{\scriptscriptstyle ZFC}} > \lambda_{\mathrm{
\scriptscriptstyle C}}|_{\scriptscriptstyle \mathrm{FC}}$, as predicted by
point \emph{b} in \sect \ref{sec:general-comparison}. At low fields, the
sequence in magnitude changes $\lambda_{\mathrm{\scriptscriptstyle
C}}|_{\mathrm{\scriptscriptstyle ZFC}} < \lambda_{\mathrm{\scriptscriptstyle
C}}|_{\mathrm{\scriptscriptstyle FC}}$ (see point \emph{c} in
\sect\ref{sec:general-comparison}) and the field-cooled Campbell length turns
hysteretic (point \emph{d} in \sect\ref{sec:general-comparison}).

Making use of the phenomenological parameters characterizing the germanide
superconductor \cite{Kim2013}, $\lambda_0 = 426~\mathrm{nm}$ and $\xi_0 =
25~\mathrm{nm}$, we find that a defect density $n_p \sim 10^{14}\,
\mathrm{cm}^{-3}$ (corresponding to a distance between defects of order $10\,
\xi_0$) provides the correct magnitude of $\lambda_{\mathrm{
\scriptscriptstyle C}}$ and is consistent with the small density condition
$n_p a_0 \xi^2 \ll 1$.

\bigskip

\section{Conclusion}

We have investigated the linear $ac$ magnetic response of type-II
superconductors in the Shubnikov phase as characterized through the Campbell
length $\lambda_{\mathrm{\scriptscriptstyle C}}$, the penetration depth of the
$ac$ signal. Starting from the microscopic theory of strong pinning, we have
shown that the Campbell length involves specific jumps in the (multi-valued)
pinning force profile corresponding to abrupt changes in the occupation of the
force branches. With this new tool at hand, we have discussed the generic
behavior of the Campbell length (i) near and away from the Labusch point
describing the onset of strong pinning, (ii) at low and high magnetic fields,
and (iii) for both field-cooled (FC) and zero-field-cooled (ZFC) state
preparations. Several new features have been observed: first, the FC and ZFC
states probe different force jumps and hence result in (possibly even
parametrically) different Campbell lengths. Second, in the critical state, a
transient initialization changes the $dc$ field by $h_{ac}$ after which the
response follows a regular $ac$ dynamics with a finite Campbell length.
Third, for the field-cooled state, we predict a possible hysteretic response
of the Campbell length upon thermal cycling. On the road towards quantitative
predictions, we have studied the scaling behavior of four types of defects
(insulating and metallic inclusions, $\delta T_{c}$- and $\delta
\ell$-pinning) and have constructed $H$-$T$ scaling diagrams for the pinning
strength $\kappa$. Finally, we have confronted our theory with experimental
data and have found good qualitative and even semi-quantitative agreement.

The framework presented here provides a quantitative relation between the
macroscopic Campbell length $\lambda_\mathrm{\scriptscriptstyle C}$ and the
underlying microscopic pinning landscape.  The power of this approach lies in
the ability to distinguish between different vortex configurations, e.g.,
field-cooled- and zero-field-cooled states or an arbitrary state `in-between',
allowing for a spectroscopic analysis of the pinscape.  While the measurement
of $\lambda_\mathrm{\scriptscriptstyle C}$ combined with theoretical insights
provides access to bulk averaged parameters of the pinscape such as the
defects' nature (insulating, metallic, $\delta T_c$- or $\delta \ell$-type),
its density $n_p$, and shape $e_p(\vec{r})$, recent experiments using scanning
STM and scanning SQUID techniques allow for space-resolved imaging of the
pinscape \cite{Timmermans2014,Embon2015}. Together, these novel techniques
provide valuable input for advanced numerical simulations of (driven) vortex
matter, e.g., based on (time-dependent) Ginzburg-Landau theory
\cite{Sadovskyy2015a,Sadovskyy2015b}. The outcome of such simulations may then
be used to better understand the signatures observed in
experiments\cite{Roditchev2015}, thus closing the loop in a fruitful
comparison of theory, experiment, and numerical simulation. In future work it
will be interesting to uncover other types of experimental signatures
providing further information on the strong pinning landscape.

\begin{acknowledgments}
We acknowledge financial support of the Swiss National Science Foundation
(SNSF) through the NCCR MaNEP. The authors are indebted to Ruslan Prozorov for
providing us with the experimental data presented in Fig.\
\ref{fig:comp-to-exp}.
\end{acknowledgments}


\begin{thebibliography}{51}%
\makeatletter
\providecommand \@ifxundefined [1]{%
 \@ifx{#1\undefined}
}%
\providecommand \@ifnum [1]{%
 \ifnum #1\expandafter \@firstoftwo
 \else \expandafter \@secondoftwo
 \fi
}%
\providecommand \@ifx [1]{%
 \ifx #1\expandafter \@firstoftwo
 \else \expandafter \@secondoftwo
 \fi
}%
\providecommand \natexlab [1]{#1}%
\providecommand \enquote  [1]{``#1''}%
\providecommand \bibnamefont  [1]{#1}%
\providecommand \bibfnamefont [1]{#1}%
\providecommand \citenamefont [1]{#1}%
\providecommand \href@noop [0]{\@secondoftwo}%
\providecommand \href [0]{\begingroup \@sanitize@url \@href}%
\providecommand \@href[1]{\@@startlink{#1}\@@href}%
\providecommand \@@href[1]{\endgroup#1\@@endlink}%
\providecommand \@sanitize@url [0]{\catcode `\\12\catcode `\$12\catcode
  `\&12\catcode `\#12\catcode `\^12\catcode `\_12\catcode `\%12\relax}%
\providecommand \@@startlink[1]{}%
\providecommand \@@endlink[0]{}%
\providecommand \url  [0]{\begingroup\@sanitize@url \@url }%
\providecommand \@url [1]{\endgroup\@href {#1}{\urlprefix }}%
\providecommand \urlprefix  [0]{URL }%
\providecommand \Eprint [0]{\href }%
\providecommand \doibase [0]{http://dx.doi.org/}%
\providecommand \selectlanguage [0]{\@gobble}%
\providecommand \bibinfo  [0]{\@secondoftwo}%
\providecommand \bibfield  [0]{\@secondoftwo}%
\providecommand \translation [1]{[#1]}%
\providecommand \BibitemOpen [0]{}%
\providecommand \bibitemStop [0]{}%
\providecommand \bibitemNoStop [0]{.\EOS\space}%
\providecommand \EOS [0]{\spacefactor3000\relax}%
\providecommand \BibitemShut  [1]{\csname bibitem#1\endcsname}%
\let\auto@bib@innerbib\@empty
\bibitem [{\citenamefont {Shubnikov}\ \emph {et~al.}(2008)\citenamefont
  {Shubnikov}, \citenamefont {Khotkevich}, \citenamefont {Shepelev},\ and\
  \citenamefont {Riabinin}}]{Shubnikov1937}%
  \BibitemOpen
  \bibfield  {author} {\bibinfo {author} {\bibfnamefont {L.~V.}\ \bibnamefont
  {Shubnikov}}, \bibinfo {author} {\bibfnamefont {V.~I.}\ \bibnamefont
  {Khotkevich}}, \bibinfo {author} {\bibfnamefont {Y.~D.}\ \bibnamefont
  {Shepelev}}, \ and\ \bibinfo {author} {\bibfnamefont {Y.~N.}\ \bibnamefont
  {Riabinin}},\ }\href@noop {} {\bibfield  {journal} {\bibinfo  {journal}
  {{[Zh. Eksp. Teor. Fiz. \textbf{7}, 221 (1937)]} Ukrainian Journal of
  Physics}\ }\textbf {\bibinfo {volume} {53}},\ \bibinfo {pages} {42} (\bibinfo
  {year} {2008})}\BibitemShut {NoStop}%
\bibitem [{\citenamefont {Abrikosov}(1957)}]{Abrikosov1957}%
  \BibitemOpen
  \bibfield  {author} {\bibinfo {author} {\bibfnamefont {A.~A.}\ \bibnamefont
  {Abrikosov}},\ }\href@noop {} {\bibfield  {journal} {\bibinfo  {journal}
  {{[Zh. Eksp. Teor. Fiz. \textbf{32}, 1442 (1957)]} JETP}\ }\textbf {\bibinfo
  {volume} {5}},\ \bibinfo {pages} {1174} (\bibinfo {year} {1957})}\BibitemShut
  {NoStop}%
\bibitem [{\citenamefont {Bardeen}\ and\ \citenamefont
  {Stephen}(1965)}]{Bardeen1965}%
  \BibitemOpen
  \bibfield  {author} {\bibinfo {author} {\bibfnamefont {J.}~\bibnamefont
  {Bardeen}}\ and\ \bibinfo {author} {\bibfnamefont {M.~J.}\ \bibnamefont
  {Stephen}},\ }\href@noop {} {\bibfield  {journal} {\bibinfo  {journal}
  {Physical Review}\ }\textbf {\bibinfo {volume} {140}},\ \bibinfo {pages}
  {A1197} (\bibinfo {year} {1965})}\BibitemShut {NoStop}%
\bibitem [{\citenamefont {Sadovskyy}\ \emph {et~al.}(2015)\citenamefont
  {Sadovskyy}, \citenamefont {Koshelev}, \citenamefont {Phillips},
  \citenamefont {Karpeev},\ and\ \citenamefont {Glatz}}]{Sadovskyy2015a}%
  \BibitemOpen
  \bibfield  {author} {\bibinfo {author} {\bibfnamefont {I.~A.}\ \bibnamefont
  {Sadovskyy}}, \bibinfo {author} {\bibfnamefont {A.~E.}\ \bibnamefont
  {Koshelev}}, \bibinfo {author} {\bibfnamefont {C.~L.}\ \bibnamefont
  {Phillips}}, \bibinfo {author} {\bibfnamefont {D.~A.}\ \bibnamefont
  {Karpeev}}, \ and\ \bibinfo {author} {\bibfnamefont {A.}~\bibnamefont
  {Glatz}},\ }\href@noop {} {\bibfield  {journal} {\bibinfo  {journal} {Journal
  of Computational Physics}\ }\textbf {\bibinfo {volume} {294}},\ \bibinfo
  {pages} {639} (\bibinfo {year} {2015})}\BibitemShut {NoStop}%
\bibitem [{\citenamefont {Sadovskyy~\emph{et al.}}(2015)}]{Sadovskyy2015b}%
  \BibitemOpen
  \bibfield  {author} {\bibinfo {author} {\bibfnamefont {I.~A.}\ \bibnamefont
  {Sadovskyy~\emph{et al.}}},\ }\href@noop {} {\bibfield  {journal} {\bibinfo
  {journal} {unpublished}\ } (\bibinfo {year} {2015})}\BibitemShut {NoStop}%
\bibitem [{\citenamefont {Campbell}(1969)}]{Campbell1969}%
  \BibitemOpen
  \bibfield  {author} {\bibinfo {author} {\bibfnamefont {A.~M.}\ \bibnamefont
  {Campbell}},\ }\href@noop {} {\bibfield  {journal} {\bibinfo  {journal}
  {Journal of Physics C: Solid State Physics}\ }\textbf {\bibinfo {volume}
  {2}},\ \bibinfo {pages} {1492} (\bibinfo {year} {1969})}\BibitemShut
  {NoStop}%
\bibitem [{\citenamefont {Campbell}(1971)}]{Campbell1971}%
  \BibitemOpen
  \bibfield  {author} {\bibinfo {author} {\bibfnamefont {A.~M.}\ \bibnamefont
  {Campbell}},\ }\href@noop {} {\bibfield  {journal} {\bibinfo  {journal}
  {Journal of Physics C: Solid State Physics}\ }\textbf {\bibinfo {volume}
  {4}},\ \bibinfo {pages} {3186} (\bibinfo {year} {1971})}\BibitemShut
  {NoStop}%
\bibitem [{\citenamefont {Lowell}(1972{\natexlab{a}})}]{Lowell1972_1}%
  \BibitemOpen
  \bibfield  {author} {\bibinfo {author} {\bibfnamefont {J.}~\bibnamefont
  {Lowell}},\ }\href@noop {} {\bibfield  {journal} {\bibinfo  {journal}
  {Journal of Physics F: Metal Physics}\ }\textbf {\bibinfo {volume} {2}},\
  \bibinfo {pages} {547} (\bibinfo {year} {1972}{\natexlab{a}})}\BibitemShut
  {NoStop}%
\bibitem [{\citenamefont {Campbell}(1978)}]{Campbell1978_1}%
  \BibitemOpen
  \bibfield  {author} {\bibinfo {author} {\bibfnamefont {A.~M.}\ \bibnamefont
  {Campbell}},\ }\href@noop {} {\bibfield  {journal} {\bibinfo  {journal}
  {Philosophical Magazine Part B}\ }\textbf {\bibinfo {volume} {37}},\ \bibinfo
  {pages} {149} (\bibinfo {year} {1978})}\BibitemShut {NoStop}%
\bibitem [{\citenamefont {Bednorz}\ and\ \citenamefont
  {{M{\"u}ller}}(1986)}]{Bednorz1986}%
  \BibitemOpen
  \bibfield  {author} {\bibinfo {author} {\bibfnamefont {J.~G.}\ \bibnamefont
  {Bednorz}}\ and\ \bibinfo {author} {\bibfnamefont {K.~A.}\ \bibnamefont
  {{M{\"u}ller}}},\ }\href@noop {} {\bibfield  {journal} {\bibinfo  {journal}
  {Zeitschrift f{\"u}r Physik B}\ }\textbf {\bibinfo {volume} {64}},\ \bibinfo
  {pages} {189} (\bibinfo {year} {1986})}\BibitemShut {NoStop}%
\bibitem [{\citenamefont {Brandt}(1991)}]{Brandt1991}%
  \BibitemOpen
  \bibfield  {author} {\bibinfo {author} {\bibfnamefont {E.~H.}\ \bibnamefont
  {Brandt}},\ }\href@noop {} {\bibfield  {journal} {\bibinfo  {journal}
  {Physical Review Letters}\ }\textbf {\bibinfo {volume} {67}},\ \bibinfo
  {pages} {2219} (\bibinfo {year} {1991})}\BibitemShut {NoStop}%
\bibitem [{\citenamefont {Koshelev}\ and\ \citenamefont
  {Vinokur}(1991)}]{Koshelev1991}%
  \BibitemOpen
  \bibfield  {author} {\bibinfo {author} {\bibfnamefont {A.~E.}\ \bibnamefont
  {Koshelev}}\ and\ \bibinfo {author} {\bibfnamefont {V.~M.}\ \bibnamefont
  {Vinokur}},\ }\href@noop {} {\bibfield  {journal} {\bibinfo  {journal}
  {Physica C}\ }\textbf {\bibinfo {volume} {173}},\ \bibinfo {pages} {465}
  (\bibinfo {year} {1991})}\BibitemShut {NoStop}%
\bibitem [{\citenamefont {Coffey}\ and\ \citenamefont
  {Clem}(1991)}]{Coffey1991}%
  \BibitemOpen
  \bibfield  {author} {\bibinfo {author} {\bibfnamefont {M.~W.}\ \bibnamefont
  {Coffey}}\ and\ \bibinfo {author} {\bibfnamefont {J.~R.}\ \bibnamefont
  {Clem}},\ }\href {\doibase 10.1103/PhysRevLett.67.386} {\bibfield  {journal}
  {\bibinfo  {journal} {Physical Review Letters}\ }\textbf {\bibinfo {volume}
  {67}},\ \bibinfo {pages} {386} (\bibinfo {year} {1991})}\BibitemShut
  {NoStop}%
\bibitem [{\citenamefont {Coffey}\ and\ \citenamefont
  {Clem}(1992)}]{Coffey1992}%
  \BibitemOpen
  \bibfield  {author} {\bibinfo {author} {\bibfnamefont {M.~W.}\ \bibnamefont
  {Coffey}}\ and\ \bibinfo {author} {\bibfnamefont {J.~R.}\ \bibnamefont
  {Clem}},\ }\href {\doibase 10.1103/PhysRevB.45.9872} {\bibfield  {journal}
  {\bibinfo  {journal} {Physical Review B}\ }\textbf {\bibinfo {volume} {45}},\
  \bibinfo {pages} {9872} (\bibinfo {year} {1992})}\BibitemShut {NoStop}%
\bibitem [{\citenamefont {van~der Beek}\ \emph {et~al.}(1993)\citenamefont
  {van~der Beek}, \citenamefont {Geshkenbein},\ and\ \citenamefont
  {Vinokur}}]{vanderBeek1993}%
  \BibitemOpen
  \bibfield  {author} {\bibinfo {author} {\bibfnamefont {C.~J.}\ \bibnamefont
  {van~der Beek}}, \bibinfo {author} {\bibfnamefont {V.~B.}\ \bibnamefont
  {Geshkenbein}}, \ and\ \bibinfo {author} {\bibfnamefont {V.~M.}\ \bibnamefont
  {Vinokur}},\ }\href@noop {} {\bibfield  {journal} {\bibinfo  {journal}
  {Physical Review B}\ }\textbf {\bibinfo {volume} {48}},\ \bibinfo {pages}
  {3393 } (\bibinfo {year} {1993})}\BibitemShut {NoStop}%
\bibitem [{\citenamefont {G{\"o}m{\"o}ry}(1997)}]{Gomory1997}%
  \BibitemOpen
  \bibfield  {author} {\bibinfo {author} {\bibfnamefont {F.}~\bibnamefont
  {G{\"o}m{\"o}ry}},\ }\href@noop {} {\bibfield  {journal} {\bibinfo  {journal}
  {Superconductor Science and Technology}\ }\textbf {\bibinfo {volume} {10}},\
  \bibinfo {pages} {523} (\bibinfo {year} {1997})}\BibitemShut {NoStop}%
\bibitem [{\citenamefont {Prozorov}\ \emph {et~al.}(2003)\citenamefont
  {Prozorov}, \citenamefont {Giannetta}, \citenamefont {Kameda}, \citenamefont
  {Tamegai}, \citenamefont {Schlueter},\ and\ \citenamefont
  {Fournier}}]{Prozorov2003}%
  \BibitemOpen
  \bibfield  {author} {\bibinfo {author} {\bibfnamefont {R.}~\bibnamefont
  {Prozorov}}, \bibinfo {author} {\bibfnamefont {R.~W.}\ \bibnamefont
  {Giannetta}}, \bibinfo {author} {\bibfnamefont {N.}~\bibnamefont {Kameda}},
  \bibinfo {author} {\bibfnamefont {T.}~\bibnamefont {Tamegai}}, \bibinfo
  {author} {\bibfnamefont {J.~A.}\ \bibnamefont {Schlueter}}, \ and\ \bibinfo
  {author} {\bibfnamefont {P.}~\bibnamefont {Fournier}},\ }\href@noop {}
  {\bibfield  {journal} {\bibinfo  {journal} {Physical Review B}\ }\textbf
  {\bibinfo {volume} {67}},\ \bibinfo {pages} {184501} (\bibinfo {year}
  {2003})}\BibitemShut {NoStop}%
\bibitem [{\citenamefont {Bean}(1962)}]{Bean1962}%
  \BibitemOpen
  \bibfield  {author} {\bibinfo {author} {\bibfnamefont {C.~P.}\ \bibnamefont
  {Bean}},\ }\href@noop {} {\bibfield  {journal} {\bibinfo  {journal} {Physical
  Review Letters}\ }\textbf {\bibinfo {volume} {8}},\ \bibinfo {pages} {250}
  (\bibinfo {year} {1962})}\BibitemShut {NoStop}%
\bibitem [{\citenamefont {Bak}\ \emph {et~al.}(1987)\citenamefont {Bak},
  \citenamefont {Tang},\ and\ \citenamefont {Wiesenfeld}}]{Bak1987}%
  \BibitemOpen
  \bibfield  {author} {\bibinfo {author} {\bibfnamefont {P.}~\bibnamefont
  {Bak}}, \bibinfo {author} {\bibfnamefont {C.}~\bibnamefont {Tang}}, \ and\
  \bibinfo {author} {\bibfnamefont {K.}~\bibnamefont {Wiesenfeld}},\
  }\href@noop {} {\bibfield  {journal} {\bibinfo  {journal} {Physical Review
  Letters}\ }\textbf {\bibinfo {volume} {59}},\ \bibinfo {pages} {381}
  (\bibinfo {year} {1987})}\BibitemShut {NoStop}%
\bibitem [{\citenamefont {Kim}\ \emph {et~al.}(2013)\citenamefont {Kim},
  \citenamefont {Sung}, \citenamefont {Cho}, \citenamefont {Tanatar},\ and\
  \citenamefont {Prozorov}}]{Kim2013}%
  \BibitemOpen
  \bibfield  {author} {\bibinfo {author} {\bibfnamefont {H.}~\bibnamefont
  {Kim}}, \bibinfo {author} {\bibfnamefont {N.~H.}\ \bibnamefont {Sung}},
  \bibinfo {author} {\bibfnamefont {B.~K.}\ \bibnamefont {Cho}}, \bibinfo
  {author} {\bibfnamefont {M.~A.}\ \bibnamefont {Tanatar}}, \ and\ \bibinfo
  {author} {\bibfnamefont {R.}~\bibnamefont {Prozorov}},\ }\href@noop {}
  {\bibfield  {journal} {\bibinfo  {journal} {Physical Review B}\ }\textbf
  {\bibinfo {volume} {87}},\ \bibinfo {pages} {094515} (\bibinfo {year}
  {2013})}\BibitemShut {NoStop}%
\bibitem [{\citenamefont {Larkin}\ and\ \citenamefont
  {Ovchinnikov}(1979)}]{Larkin1979}%
  \BibitemOpen
  \bibfield  {author} {\bibinfo {author} {\bibfnamefont {A.~I.}\ \bibnamefont
  {Larkin}}\ and\ \bibinfo {author} {\bibfnamefont {Y.~N.}\ \bibnamefont
  {Ovchinnikov}},\ }\href@noop {} {\bibfield  {journal} {\bibinfo  {journal}
  {Journal of Low Temperature Physics}\ }\textbf {\bibinfo {volume} {34}},\
  \bibinfo {pages} {409} (\bibinfo {year} {1979})}\BibitemShut {NoStop}%
\bibitem [{\citenamefont {Blatter}\ \emph {et~al.}(2004)\citenamefont
  {Blatter}, \citenamefont {Geshkenbein},\ and\ \citenamefont
  {Koopmann}}]{Blatter2004}%
  \BibitemOpen
  \bibfield  {author} {\bibinfo {author} {\bibfnamefont {G.}~\bibnamefont
  {Blatter}}, \bibinfo {author} {\bibfnamefont {V.~B.}\ \bibnamefont
  {Geshkenbein}}, \ and\ \bibinfo {author} {\bibfnamefont {J.~A.~G.}\
  \bibnamefont {Koopmann}},\ }\href@noop {} {\bibfield  {journal} {\bibinfo
  {journal} {Physical Review Letters}\ }\textbf {\bibinfo {volume} {92}},\
  \bibinfo {pages} {067009} (\bibinfo {year} {2004})}\BibitemShut {NoStop}%
\bibitem [{\citenamefont {Timmermans}\ \emph {et~al.}(2014)\citenamefont
  {Timmermans}, \citenamefont {Samuely}, \citenamefont {Raes}, \citenamefont
  {de~Vondel},\ and\ \citenamefont {Moshchalkov}}]{Timmermans2014}%
  \BibitemOpen
  \bibfield  {author} {\bibinfo {author} {\bibfnamefont {M.}~\bibnamefont
  {Timmermans}}, \bibinfo {author} {\bibfnamefont {T.}~\bibnamefont {Samuely}},
  \bibinfo {author} {\bibfnamefont {B.}~\bibnamefont {Raes}}, \bibinfo {author}
  {\bibfnamefont {J.~V.}\ \bibnamefont {de~Vondel}}, \ and\ \bibinfo {author}
  {\bibfnamefont {V.~V.}\ \bibnamefont {Moshchalkov}},\ }\href@noop {}
  {\bibfield  {journal} {\bibinfo  {journal} {ACS Nano}\ }\textbf {\bibinfo
  {volume} {8}},\ \bibinfo {pages} {2782} (\bibinfo {year} {2014})}\BibitemShut
  {NoStop}%
\bibitem [{\citenamefont {Embon}\ \emph {et~al.}(2015)\citenamefont {Embon},
  \citenamefont {Anahory}, \citenamefont {Suhov}, \citenamefont {Halbertal},
  \citenamefont {Cuppens}, \citenamefont {Yakovenko}, \citenamefont {Uri},
  \citenamefont {Myasoedov}, \citenamefont {Rappaport}, \citenamefont {Huber},
  \citenamefont {Gurevich},\ and\ \citenamefont {Zeldov}}]{Embon2015}%
  \BibitemOpen
  \bibfield  {author} {\bibinfo {author} {\bibfnamefont {L.}~\bibnamefont
  {Embon}}, \bibinfo {author} {\bibfnamefont {Y.}~\bibnamefont {Anahory}},
  \bibinfo {author} {\bibfnamefont {A.}~\bibnamefont {Suhov}}, \bibinfo
  {author} {\bibfnamefont {D.}~\bibnamefont {Halbertal}}, \bibinfo {author}
  {\bibfnamefont {J.}~\bibnamefont {Cuppens}}, \bibinfo {author} {\bibfnamefont
  {A.}~\bibnamefont {Yakovenko}}, \bibinfo {author} {\bibfnamefont
  {A.}~\bibnamefont {Uri}}, \bibinfo {author} {\bibfnamefont {Y.}~\bibnamefont
  {Myasoedov}}, \bibinfo {author} {\bibfnamefont {M.~L.}\ \bibnamefont
  {Rappaport}}, \bibinfo {author} {\bibfnamefont {M.~E.}\ \bibnamefont
  {Huber}}, \bibinfo {author} {\bibfnamefont {A.}~\bibnamefont {Gurevich}}, \
  and\ \bibinfo {author} {\bibfnamefont {E.}~\bibnamefont {Zeldov}},\
  }\href@noop {} {\bibfield  {journal} {\bibinfo  {journal} {Scientific
  Reports}\ }\textbf {\bibinfo {volume} {5}},\ \bibinfo {pages} {7598}
  (\bibinfo {year} {2015})}\BibitemShut {NoStop}%
\bibitem [{\citenamefont {Willa}\ \emph
  {et~al.}(2015{\natexlab{a}})\citenamefont {Willa}, \citenamefont
  {Geshkenbein}, \citenamefont {Prozorov},\ and\ \citenamefont
  {Blatter}}]{Willa2015a}%
  \BibitemOpen
  \bibfield  {author} {\bibinfo {author} {\bibfnamefont {R.}~\bibnamefont
  {Willa}}, \bibinfo {author} {\bibfnamefont {V.~B.}\ \bibnamefont
  {Geshkenbein}}, \bibinfo {author} {\bibfnamefont {R.}~\bibnamefont
  {Prozorov}}, \ and\ \bibinfo {author} {\bibfnamefont {G.}~\bibnamefont
  {Blatter}},\ }\href@noop {} {\bibfield  {journal} {\bibinfo  {journal}
  {Physical Review Letters}\ }\textbf {\bibinfo {volume} {115}},\ \bibinfo
  {pages} {207001} (\bibinfo {year} {2015}{\natexlab{a}})}\BibitemShut
  {NoStop}%
\bibitem [{\citenamefont {Willa}\ \emph
  {et~al.}(2015{\natexlab{b}})\citenamefont {Willa}, \citenamefont
  {Geshkenbein},\ and\ \citenamefont {Blatter}}]{Willa2015b}%
  \BibitemOpen
  \bibfield  {author} {\bibinfo {author} {\bibfnamefont {R.}~\bibnamefont
  {Willa}}, \bibinfo {author} {\bibfnamefont {V.~B.}\ \bibnamefont
  {Geshkenbein}}, \ and\ \bibinfo {author} {\bibfnamefont {G.}~\bibnamefont
  {Blatter}},\ }\href@noop {} {\bibfield  {journal} {\bibinfo  {journal}
  {Physical Review B}\ }\textbf {\bibinfo {volume} {92}},\ \bibinfo {pages}
  {134501} (\bibinfo {year} {2015}{\natexlab{b}})}\BibitemShut {NoStop}%
\bibitem [{\citenamefont {Labusch}(1969)}]{Labusch1969}%
  \BibitemOpen
  \bibfield  {author} {\bibinfo {author} {\bibfnamefont {R.}~\bibnamefont
  {Labusch}},\ }\href@noop {} {\bibfield  {journal} {\bibinfo  {journal}
  {Crystal Lattice Defects}\ }\textbf {\bibinfo {volume} {1}},\ \bibinfo
  {pages} {1} (\bibinfo {year} {1969})}\BibitemShut {NoStop}%
\bibitem [{\citenamefont {Zeldov}\ \emph {et~al.}(1994)\citenamefont {Zeldov},
  \citenamefont {Larkin}, \citenamefont {Geshkenbein}, \citenamefont
  {Konczykowski}, \citenamefont {Majer}, \citenamefont {Khaykovich},
  \citenamefont {Vinokur},\ and\ \citenamefont {Shtrikman}}]{Zeldov1994}%
  \BibitemOpen
  \bibfield  {author} {\bibinfo {author} {\bibfnamefont {E.}~\bibnamefont
  {Zeldov}}, \bibinfo {author} {\bibfnamefont {A.~I.}\ \bibnamefont {Larkin}},
  \bibinfo {author} {\bibfnamefont {V.~B.}\ \bibnamefont {Geshkenbein}},
  \bibinfo {author} {\bibfnamefont {M.}~\bibnamefont {Konczykowski}}, \bibinfo
  {author} {\bibfnamefont {D.}~\bibnamefont {Majer}}, \bibinfo {author}
  {\bibfnamefont {B.}~\bibnamefont {Khaykovich}}, \bibinfo {author}
  {\bibfnamefont {V.~M.}\ \bibnamefont {Vinokur}}, \ and\ \bibinfo {author}
  {\bibfnamefont {H.}~\bibnamefont {Shtrikman}},\ }\href@noop {} {\bibfield
  {journal} {\bibinfo  {journal} {Physical Review Letters}\ }\textbf {\bibinfo
  {volume} {73}},\ \bibinfo {pages} {1428} (\bibinfo {year}
  {1994})}\BibitemShut {NoStop}%
\bibitem [{\citenamefont {Willa}\ \emph {et~al.}(2014)\citenamefont {Willa},
  \citenamefont {Geshkenbein},\ and\ \citenamefont {Blatter}}]{Willa2014}%
  \BibitemOpen
  \bibfield  {author} {\bibinfo {author} {\bibfnamefont {R.}~\bibnamefont
  {Willa}}, \bibinfo {author} {\bibfnamefont {V.~B.}\ \bibnamefont
  {Geshkenbein}}, \ and\ \bibinfo {author} {\bibfnamefont {G.}~\bibnamefont
  {Blatter}},\ }\href@noop {} {\bibfield  {journal} {\bibinfo  {journal}
  {Physical Review B}\ }\textbf {\bibinfo {volume} {89}},\ \bibinfo {pages}
  {104514} (\bibinfo {year} {2014})}\BibitemShut {NoStop}%
\bibitem [{\citenamefont {Blatter}\ \emph {et~al.}(1994)\citenamefont
  {Blatter}, \citenamefont {Feigel'man}, \citenamefont {Geshkenbein},
  \citenamefont {Larkin},\ and\ \citenamefont {Vinokur}}]{Blatter1994}%
  \BibitemOpen
  \bibfield  {author} {\bibinfo {author} {\bibfnamefont {G.}~\bibnamefont
  {Blatter}}, \bibinfo {author} {\bibfnamefont {M.~V.}\ \bibnamefont
  {Feigel'man}}, \bibinfo {author} {\bibfnamefont {V.~B.}\ \bibnamefont
  {Geshkenbein}}, \bibinfo {author} {\bibfnamefont {A.~I.}\ \bibnamefont
  {Larkin}}, \ and\ \bibinfo {author} {\bibfnamefont {V.~M.}\ \bibnamefont
  {Vinokur}},\ }\href@noop {} {\bibfield  {journal} {\bibinfo  {journal}
  {Review of Modern Physics}\ }\textbf {\bibinfo {volume} {66}},\ \bibinfo
  {pages} {1125} (\bibinfo {year} {1994})}\BibitemShut {NoStop}%
\bibitem [{\citenamefont {Brandt}(1977{\natexlab{a}})}]{Brandt1977a}%
  \BibitemOpen
  \bibfield  {author} {\bibinfo {author} {\bibfnamefont {E.~H.}\ \bibnamefont
  {Brandt}},\ }\href@noop {} {\bibfield  {journal} {\bibinfo  {journal}
  {Journal of Low Temperature Physics}\ }\textbf {\bibinfo {volume} {26}},\
  \bibinfo {pages} {709} (\bibinfo {year} {1977}{\natexlab{a}})}\BibitemShut
  {NoStop}%
\bibitem [{\citenamefont {Brandt}(1977{\natexlab{b}})}]{Brandt1977b}%
  \BibitemOpen
  \bibfield  {author} {\bibinfo {author} {\bibfnamefont {E.~H.}\ \bibnamefont
  {Brandt}},\ }\href@noop {} {\bibfield  {journal} {\bibinfo  {journal}
  {Journal of Low Temperature Physics}\ }\textbf {\bibinfo {volume} {26}},\
  \bibinfo {pages} {735} (\bibinfo {year} {1977}{\natexlab{b}})}\BibitemShut
  {NoStop}%
\bibitem [{\citenamefont {Blatter}\ and\ \citenamefont
  {Geshkenbein}(2008)}]{Blatter2008}%
  \BibitemOpen
  \bibfield  {author} {\bibinfo {author} {\bibfnamefont {G.}~\bibnamefont
  {Blatter}}\ and\ \bibinfo {author} {\bibfnamefont {V.~B.}\ \bibnamefont
  {Geshkenbein}},\ }in\ \href@noop {} {\emph {\bibinfo {booktitle}
  {Superconductivity}}},\ \bibinfo {editor} {edited by\ \bibinfo {editor}
  {\bibfnamefont {K.}~\bibnamefont {Bennemann}}\ and\ \bibinfo {editor}
  {\bibfnamefont {J.}~\bibnamefont {Ketterson}}}\ (\bibinfo  {publisher}
  {Springer Berlin Heidelberg},\ \bibinfo {year} {2008})\ pp.\ \bibinfo {pages}
  {495--637}\BibitemShut {NoStop}%
\bibitem [{\citenamefont {Koopmann}\ \emph {et~al.}(2004)\citenamefont
  {Koopmann}, \citenamefont {Geshkenbein},\ and\ \citenamefont
  {Blatter}}]{Koopmann2004}%
  \BibitemOpen
  \bibfield  {author} {\bibinfo {author} {\bibfnamefont {J.~A.~G.}\
  \bibnamefont {Koopmann}}, \bibinfo {author} {\bibfnamefont {V.~B.}\
  \bibnamefont {Geshkenbein}}, \ and\ \bibinfo {author} {\bibfnamefont
  {G.}~\bibnamefont {Blatter}},\ }\href@noop {} {\bibfield  {journal} {\bibinfo
   {journal} {Physica C}\ }\textbf {\bibinfo {volume} {404}},\ \bibinfo {pages}
  {209} (\bibinfo {year} {2004})}\BibitemShut {NoStop}%
\bibitem [{\citenamefont {Ovchinnikov}\ and\ \citenamefont
  {Ivlev}(1991)}]{Ovchinnikov1991}%
  \BibitemOpen
  \bibfield  {author} {\bibinfo {author} {\bibfnamefont {Y.~N.}\ \bibnamefont
  {Ovchinnikov}}\ and\ \bibinfo {author} {\bibfnamefont {B.~I.}\ \bibnamefont
  {Ivlev}},\ }\href {\doibase 10.1103/PhysRevB.43.8024} {\bibfield  {journal}
  {\bibinfo  {journal} {Physical Review B}\ }\textbf {\bibinfo {volume} {43}},\
  \bibinfo {pages} {8024} (\bibinfo {year} {1991})}\BibitemShut {NoStop}%
\bibitem [{\citenamefont {Sch\"onenberger}\ \emph {et~al.}(1996)\citenamefont
  {Sch\"onenberger}, \citenamefont {Larkin}, \citenamefont {Heeb},
  \citenamefont {Geshkenbein},\ and\ \citenamefont
  {Blatter}}]{Schoenenberger1996}%
  \BibitemOpen
  \bibfield  {author} {\bibinfo {author} {\bibfnamefont {A.}~\bibnamefont
  {Sch\"onenberger}}, \bibinfo {author} {\bibfnamefont {A.~I.}\ \bibnamefont
  {Larkin}}, \bibinfo {author} {\bibfnamefont {E.}~\bibnamefont {Heeb}},
  \bibinfo {author} {\bibfnamefont {V.~B.}\ \bibnamefont {Geshkenbein}}, \ and\
  \bibinfo {author} {\bibfnamefont {G.}~\bibnamefont {Blatter}},\ }\href@noop
  {} {\bibfield  {journal} {\bibinfo  {journal} {Physical Review Letters}\
  }\textbf {\bibinfo {volume} {77}},\ \bibinfo {pages} {4636} (\bibinfo {year}
  {1996})}\BibitemShut {NoStop}%
\bibitem [{\citenamefont {Koshelev}()}]{KoshelevPrivComm}%
  \BibitemOpen
  \bibfield  {author} {\bibinfo {author} {\bibfnamefont {A.~E.}\ \bibnamefont
  {Koshelev}},\ }\href@noop {} {}\bibinfo {howpublished} {private
  communication}\BibitemShut {NoStop}%
\bibitem [{\citenamefont {Brandt}(1986)}]{Brandt1986}%
  \BibitemOpen
  \bibfield  {author} {\bibinfo {author} {\bibfnamefont {E.~H.}\ \bibnamefont
  {Brandt}},\ }\href@noop {} {\bibfield  {journal} {\bibinfo  {journal}
  {Physical Review B}\ }\textbf {\bibinfo {volume} {34}},\ \bibinfo {pages}
  {6514} (\bibinfo {year} {1986})}\BibitemShut {NoStop}%
\bibitem [{\citenamefont {Ginzburg}\ and\ \citenamefont
  {Landau}(1950)}]{Ginzburg1950}%
  \BibitemOpen
  \bibfield  {author} {\bibinfo {author} {\bibfnamefont {V.~L.}\ \bibnamefont
  {Ginzburg}}\ and\ \bibinfo {author} {\bibfnamefont {L.~D.}\ \bibnamefont
  {Landau}},\ }\href@noop {} {\bibfield  {journal} {\bibinfo  {journal} {{[L.
  D. Landau, \emph{Collected papers} p. 546 (1965), Pergamon Press, Oxford]}
  Zh. Eksp. Teor. Fiz.}\ }\textbf {\bibinfo {volume} {20}},\ \bibinfo {pages}
  {1064} (\bibinfo {year} {1950})}\BibitemShut {NoStop}%
\bibitem [{\citenamefont {Schmid}(1966)}]{Schmid1966}%
  \BibitemOpen
  \bibfield  {author} {\bibinfo {author} {\bibfnamefont {A.}~\bibnamefont
  {Schmid}},\ }\href@noop {} {\bibfield  {journal} {\bibinfo  {journal} {Physik
  der kondensierten Materie}\ }\textbf {\bibinfo {volume} {5}},\ \bibinfo
  {pages} {302} (\bibinfo {year} {1966})}\BibitemShut {NoStop}%
\bibitem [{\citenamefont {Clem}(1975)}]{Clem1975}%
  \BibitemOpen
  \bibfield  {author} {\bibinfo {author} {\bibfnamefont {J.~R.}\ \bibnamefont
  {Clem}},\ }\href@noop {} {\bibfield  {journal} {\bibinfo  {journal} {Journal
  of Low Temperature Physics}\ }\textbf {\bibinfo {volume} {18}},\ \bibinfo
  {pages} {427} (\bibinfo {year} {1975})}\BibitemShut {NoStop}%
\bibitem [{\citenamefont {Saint-James}\ \emph {et~al.}(1969)\citenamefont
  {Saint-James}, \citenamefont {Sarma},\ and\ \citenamefont
  {Thomas}}]{SaintJames1969}%
  \BibitemOpen
  \bibfield  {author} {\bibinfo {author} {\bibfnamefont {D.}~\bibnamefont
  {Saint-James}}, \bibinfo {author} {\bibfnamefont {G.}~\bibnamefont {Sarma}},
  \ and\ \bibinfo {author} {\bibfnamefont {E.~J.}\ \bibnamefont {Thomas}},\
  }\href@noop {} {\emph {\bibinfo {title} {{Type II Superconductivity}}}},\
  \bibinfo {series} {International series of monographs in natural philosophy},
  Vol.~\bibinfo {volume} {17}\ (\bibinfo  {publisher} {Oxford: Pergamon},\
  \bibinfo {year} {1969})\BibitemShut {NoStop}%
\bibitem [{\citenamefont {Ruggeri}\ and\ \citenamefont
  {Thouless}(1976)}]{Ruggeri1976}%
  \BibitemOpen
  \bibfield  {author} {\bibinfo {author} {\bibfnamefont {G.~J.}\ \bibnamefont
  {Ruggeri}}\ and\ \bibinfo {author} {\bibfnamefont {D.~J.}\ \bibnamefont
  {Thouless}},\ }\href@noop {} {\bibfield  {journal} {\bibinfo  {journal}
  {Journal of Physics F: Metal Physics}\ }\textbf {\bibinfo {volume} {6}},\
  \bibinfo {pages} {2063} (\bibinfo {year} {1976})}\BibitemShut {NoStop}%
\bibitem [{\citenamefont {DeGennes}(1966)}]{DeGennes1966}%
  \BibitemOpen
  \bibfield  {author} {\bibinfo {author} {\bibfnamefont {P.~G.}\ \bibnamefont
  {DeGennes}},\ }\href@noop {} {\emph {\bibinfo {title} {{Superconductivity of
  Metals and Alloys}}}}\ (\bibinfo  {publisher} {West View Press},\ \bibinfo
  {year} {1966})\BibitemShut {NoStop}%
\bibitem [{\citenamefont {Autler}\ \emph {et~al.}(1962)\citenamefont {Autler},
  \citenamefont {Rosenblum},\ and\ \citenamefont {Gooen}}]{Autler1962}%
  \BibitemOpen
  \bibfield  {author} {\bibinfo {author} {\bibfnamefont {S.~H.}\ \bibnamefont
  {Autler}}, \bibinfo {author} {\bibfnamefont {E.~S.}\ \bibnamefont
  {Rosenblum}}, \ and\ \bibinfo {author} {\bibfnamefont {K.~H.}\ \bibnamefont
  {Gooen}},\ }\href@noop {} {\bibfield  {journal} {\bibinfo  {journal}
  {Physical Review Letters}\ }\textbf {\bibinfo {volume} {9}},\ \bibinfo
  {pages} {489} (\bibinfo {year} {1962})}\BibitemShut {NoStop}%
\bibitem [{\citenamefont {DeSorbo}(1964)}]{DeSorbo1964}%
  \BibitemOpen
  \bibfield  {author} {\bibinfo {author} {\bibfnamefont {W.}~\bibnamefont
  {DeSorbo}},\ }\href@noop {} {\bibfield  {journal} {\bibinfo  {journal}
  {Review of Modern Physics}\ }\textbf {\bibinfo {volume} {36}},\ \bibinfo
  {pages} {90} (\bibinfo {year} {1964})}\BibitemShut {NoStop}%
\bibitem [{\citenamefont {Gorkov}(1959)}]{Gorkov1959}%
  \BibitemOpen
  \bibfield  {author} {\bibinfo {author} {\bibfnamefont {L.~P.}\ \bibnamefont
  {Gorkov}},\ }\href@noop {} {\bibfield  {journal} {\bibinfo  {journal} {{[Zh.
  Eksp. Teor. Fiz. \textbf{36}, 1918 (1959)]} JETP}\ }\textbf {\bibinfo
  {volume} {9}},\ \bibinfo {pages} {1364} (\bibinfo {year} {1959})}\BibitemShut
  {NoStop}%
\bibitem [{\citenamefont {Lowell}(1972{\natexlab{b}})}]{Lowell1972_2}%
  \BibitemOpen
  \bibfield  {author} {\bibinfo {author} {\bibfnamefont {J.}~\bibnamefont
  {Lowell}},\ }\href@noop {} {\bibfield  {journal} {\bibinfo  {journal}
  {Journal of Physics F: Metal Physics}\ }\textbf {\bibinfo {volume} {2}},\
  \bibinfo {pages} {559} (\bibinfo {year} {1972}{\natexlab{b}})}\BibitemShut
  {NoStop}%
\bibitem [{\citenamefont {Gordon}\ \emph {et~al.}(2013)\citenamefont {Gordon},
  \citenamefont {Zhigadlo}, \citenamefont {Weyeneth}, \citenamefont {Katrych},\
  and\ \citenamefont {Prozorov}}]{Gordon2013}%
  \BibitemOpen
  \bibfield  {author} {\bibinfo {author} {\bibfnamefont {R.~T.}\ \bibnamefont
  {Gordon}}, \bibinfo {author} {\bibfnamefont {N.~D.}\ \bibnamefont
  {Zhigadlo}}, \bibinfo {author} {\bibfnamefont {S.}~\bibnamefont {Weyeneth}},
  \bibinfo {author} {\bibfnamefont {S.}~\bibnamefont {Katrych}}, \ and\
  \bibinfo {author} {\bibfnamefont {R.}~\bibnamefont {Prozorov}},\ }\href@noop
  {} {\bibfield  {journal} {\bibinfo  {journal} {Physical Review B}\ }\textbf
  {\bibinfo {volume} {87}},\ \bibinfo {pages} {094520} (\bibinfo {year}
  {2013})}\BibitemShut {NoStop}%
\bibitem [{\citenamefont {Sung}\ \emph {et~al.}(2012)\citenamefont {Sung},
  \citenamefont {Jo},\ and\ \citenamefont {Cho}}]{Sung2012}%
  \BibitemOpen
  \bibfield  {author} {\bibinfo {author} {\bibfnamefont {N.~H.}\ \bibnamefont
  {Sung}}, \bibinfo {author} {\bibfnamefont {Y.~J.}\ \bibnamefont {Jo}}, \ and\
  \bibinfo {author} {\bibfnamefont {B.~K.}\ \bibnamefont {Cho}},\ }\href@noop
  {} {\bibfield  {journal} {\bibinfo  {journal} {Superconductor Science and
  Technology}\ }\textbf {\bibinfo {volume} {25}},\ \bibinfo {pages} {075002}
  (\bibinfo {year} {2012})}\BibitemShut {NoStop}%
\bibitem [{\citenamefont {Roditchev}\ \emph {et~al.}(2015)\citenamefont
  {Roditchev}, \citenamefont {Brun}, \citenamefont {Serrier-Garcia},
  \citenamefont {Cuevas}, \citenamefont {Bessa}, \citenamefont {Milosevic},
  \citenamefont {Debontridder}, \citenamefont {Stolyarov},\ and\ \citenamefont
  {Cren}}]{Roditchev2015}%
  \BibitemOpen
  \bibfield  {author} {\bibinfo {author} {\bibfnamefont {D.}~\bibnamefont
  {Roditchev}}, \bibinfo {author} {\bibfnamefont {C.}~\bibnamefont {Brun}},
  \bibinfo {author} {\bibfnamefont {L.}~\bibnamefont {Serrier-Garcia}},
  \bibinfo {author} {\bibfnamefont {J.~C.}\ \bibnamefont {Cuevas}}, \bibinfo
  {author} {\bibfnamefont {V.~H.~L.}\ \bibnamefont {Bessa}}, \bibinfo {author}
  {\bibfnamefont {M.~V.}\ \bibnamefont {Milosevic}}, \bibinfo {author}
  {\bibfnamefont {F.}~\bibnamefont {Debontridder}}, \bibinfo {author}
  {\bibfnamefont {V.}~\bibnamefont {Stolyarov}}, \ and\ \bibinfo {author}
  {\bibfnamefont {T.}~\bibnamefont {Cren}},\ }\href@noop {} {\bibfield
  {journal} {\bibinfo  {journal} {Nature Physics}\ }\textbf {\bibinfo {volume}
  {11}},\ \bibinfo {pages} {332} (\bibinfo {year} {2015})}\BibitemShut
  {NoStop}%
\end{thebibliography}

%

\end{document}